\documentclass[pra,twocolumn,floatfix]{revtex4}
\usepackage{graphicx}
\usepackage{color}
\usepackage{amsmath}
\DeclareUnicodeCharacter{2009}{\,}
  
\begin{document}

\title{Rotating quantum droplets confined in an anharmonic potential}

\author{S. Nikolaou$^1$, G. M. Kavoulakis$^{1,2}$, and M. \"{O}gren$^{2,3}$}
\affiliation{$^1$Hellenic Mediterranean University, P.O. Box 1939, GR-71004, Heraklion, Greece
\\
$^2$HMU Research Center, Institute of Emerging Technologies, GR-71004, Heraklion, 
Greece
\\
$^3$School of Science and Technology, \"{O}rebro University, 70182 \"{O}rebro, Sweden}
\date{\today}

\begin{abstract}

We investigate the rotational properties of quantum droplets, which form in a mixture 
of two Bose-Einstein condensates, in the presence of an anharmonic trapping potential. 
We identify various phases as the atom number and the angular momentum/angular velocity 
of the trap vary. These phases include center-of-mass-like excitation (without, or 
with vortices), vortices of single and multiple quantization, etc. Finally, we compare 
our results with those of the single-component problem.

\end{abstract}

\maketitle

\section{Introduction}

It is well-known that when a superfluid rotates, fascinating effects arise,
which constitute the collection of phenomena that we call ``superfluidity"
\cite{Leggett}. The initial studies of superfluidity were focused on homogeneous 
superfluids, as, e.g., in liquid Helium which is confined in a bucket. The great 
progress that has been achieved in the field of trapped, atomic superfluids in 
the last 25 years has introduced another very important aspect in this problem. 
This is the effect of the trapping potential on the rotational response of these 
superfluid systems. 

In the initial experiments the trapping potential was harmonic. We stress that 
the harmonic potential is very special for the simple reason that the centrifugal 
potential resembles a repulsive harmonic potential, since it scales also quadratically 
with the distance from the center of the trap. As a result, when the rotational 
frequency of the trap $\Omega$ is equal to the trap frequency $\omega$, the effective 
potential that results from the sum of the two vanishes. Therefore, when the effective  
interaction between the atoms is repulsive, the atoms then fly apart. In other words,
in a purely harmonic trapping potential, $\Omega$ is limited by $\omega$. Interestingly
enough, as $\Omega$ approaches $\omega$ from below, the gas enters a highly correlated 
regime. This is a very interesting problem, which has attracted a lot of attention, 
see, e.g., the review article of Ref.\,\cite{Cooper}.

Eventually other forms of confining potentials were developed and studied, with 
the most common one being the anharmonic, quartic, potential. Such a potential was
studied both experimentally, see, e.g., Ref.\,\cite{Dalibard}, as well as theoretically, 
see, e.g., Refs.\,\cite{qq1, qq2, qq3, qq4, qq5, qq6, qq7, qq8, qq9}. Contrary to the 
case of a harmonic confining potential, in this case there is no bound in the 
value of $\Omega$. The study of this problem has shown that there is a wide variety 
of phases, which include vortices of single and multiple quantization, a vortex
lattice with, or without a hole, etc.
 
A recent and interesting development in the field of atomic superfluids has to do 
with the so-called quantum droplets, whose existence was proposed by Petrov 
\cite{Petrov}. These highly-quantum objects form in binary mixtures of Bose-Einstein 
condensed atoms. The basic idea which leads to the formation of quantum droplets is 
that by tuning the inter- and intra-atomic interaction strengths, the mean-field 
interaction energy becomes comparable with the next-order correction of the energy 
\cite{LHY}, which is essentially negligible in a single-component system (due to 
the assumption that we deal with dilute gases). Then, the balance between the mean-field 
energy and the beyond-mean-field correction to the energy gives rise to 
self-bound quantum droplets. 

This problem has attracted a lot of attention lately, see, e.g., the review articles 
\cite{rrev1, rrev2}, and Refs.\,\cite{PA, th0, th1, th2, th3, th4, th5, th6, th7, th8, 
th9, th10, th11, th12, th13, th14, th15, th16, EK, th166, add1, NKO, add2, add3, add4, 
add5}. Quantum droplets have also been observed experimentally both in mixtures of 
Bose-Einstein condensed gases \cite{qd7, qd8, qd8a, gd8b, qd8c} and in single-component 
gases with strong dipolar interactions \cite{qd1, qd2, qd3, qd4, qd5, qd6}. 

Being self-bound, quantum droplets exist in free space and do not require the 
presence of any trapping potential. On the other hand, it is both experimentally, 
as well as theoretically, very interesting to investigate the rotational response 
of this new superfluid system, in the presence of an external trapping potential. 

Motivated by the remarks of the previous paragraphs, we investigate in the present
study the rotational response of a quantum droplet under the action of an anharmonic 
potential \cite{rapanh, dynamics}. The results of Ref.\,\cite{rapanh}, which has 
studied the same problem, are consistent with the ones presented below. On the other 
hand, our study demonstrates the very rich structure of this problem, since we have 
identified numerous novel phases. In addition, Ref.\,\cite{dynamics} has studied the 
same problem, as well as the dynamics of a quantum droplet which is confined in an 
anharmonic potential. 

An interesting aspect of our study arises from the comparison of the present problem 
-- i.e., the rotational response of an anharmonically-trapped quantum droplet -- with 
that of a single-component condensate, which is confined in the same potential. For 
this problem we refer to, e.g., Refs.\,\cite{gv1,gv2,gv3,gv4,gv5,gv6,gv7,gv8,gv9,gv10,
gv11} for a single-component condensate, as well as to Ref.\,\cite{Elife} for the case 
of a binary mixture (but not in the limit where droplets form). 

As we analyse below, there are some similarities, but also some serious differences 
between the two problems. One major difference is that, while in the single-component 
problem there is an unstable phase (when the effective interaction is attractive), in 
the case of droplets such a phase is never present. In addition, in the problem of a 
single component, for a sufficiently small atom number, vortices of multiple quantization 
are always energetically favourable, independently of the sign of the effective 
interaction. On the other hand, in droplets, for a small atom number the motion 
resembles center-of-mass excitation, provided that the absolute value of the energy 
due to the nonlinear term is much larger than the energy due to the anharmonic 
potential.

According to the results that we present below, there is a wide variety of phases
in the problem of an anharmonically-confined rotating droplet. These include 
center-of-mass-like excitation, with a density distribution which varies from being 
almost axially-symmetric, up to being largely distorted. In addition, we have found 
phases with vortices of single and multiple quantization, as well as a ``mixed" phase,
which is, approximately, a combination of center-of-mass and vortex excitation. 

In what follows below we present in Sec.\,II the model that we use. We choose
to work with a fixed total angular momentum $L$, minimizing the energy for some 
fixed $L$, since this makes the problem more transparent. In Sec.\,III we present 
the results of our study, for some representative values of the atom number 
of the droplet $N$ and of the angular momentum per particle $\ell = L/N$. We 
identify the various phases which come out from our analysis. In the same section 
we also derive the function $\ell = \ell(\Omega)$, for the case where, instead of 
$\ell$, the rotational angular velocity of the trap $\Omega$ is fixed. In Sec.\,IV 
we present the general picture that results from our analysis and derive an 
experimentally-relevant phase diagram. In Sec.\,V we investigate the experimental 
relevance of our results, giving some typical values of the various parameters. 
Finally, in Sec.\,VI we summarize the main results of our study.

\section{Model}

Assuming that there is a very tight confining potential along the axis of rotation, 
we consider motion of the atoms in the perpendicular plane, i.e., two-dimensional 
motion. We also assume that the quantum droplet is confined in a two-dimensional
anharmonic potential,
\begin{equation}
  V(\rho) = \frac M 2 \omega^2 \rho^2 \left( 1 + \lambda \frac {\rho^2} {a_0^2} \right).
\end{equation} 
Here $\rho$ is the radial coordinate in cylindrical-polar coordinates, $M$ is 
the atom mass, which is assumed to be the same for the two components, $\omega$ 
is the frequency of the harmonic potential, $a_0 = \sqrt{\hbar/(M \omega)}$ is 
the oscillator length, and $\lambda$ is a (dimensionless) parameter which controls 
the ``strength" of the anharmonic part of the trapping potential. 

We consider the ``symmetric" case, where we have equal populations of atoms $N/2$ 
in the two components, equal masses, while the couplings between the same components 
are also assumed to be equal. In this case the order parameter of the two components 
$\Psi_{\uparrow}$ and $\Psi_{\downarrow}$ are equal to each other, $\Psi_{\uparrow} = 
\Psi_{\downarrow}$.

We introduce $\Psi = {\sqrt 2} \Psi_{\uparrow} = {\sqrt 2} \Psi_{\downarrow}$, and
also the unit of density
\begin{eqnarray}
  \Psi_0^2 = \frac {e^{-2 \gamma - 1}} {\pi} \frac {\ln(a_{\uparrow \downarrow}/a)} 
   {a a_{\uparrow \downarrow}}.
\end{eqnarray}
Here $a$ and $a_{\uparrow \downarrow}$ are the two-dimensional scattering lengths 
for elastic atom-atom collisions between the same species (assumed to be equal for 
the two components) and for different species, respectively, while $\gamma$ is Euler's 
constant, $\gamma \approx 0.5772$. Also \cite{PA},
\begin{eqnarray}
    \ln (a_{\uparrow \downarrow}/a) = \sqrt{\frac {\pi} 2} 
    \left( \frac {a_z} {a^{\rm 3D}} - \frac {a_z} {a_{\uparrow \downarrow}^{\rm 3D}} \right).
\end{eqnarray}
Here $a_z$ is the ``width" of the droplet along the axis of rotation, and $a^{\rm 3D}$, 
$a_{\uparrow \downarrow}^{\rm 3D}$ are the three-dimensional scattering lengths for  
elastic atom-atom collisions between the same and different species, respectively.
The unit of length that we adopt is
\begin{equation}
 x_0 = \sqrt{\frac {a a_{\uparrow \downarrow} \ln(a_{\uparrow \downarrow}/a)} 
 {4 e^{-2 \gamma - 1}}},
\end{equation}
while that of the energy, $E_0$, and of the frequency, $\omega_0$, are  
\begin{equation}
  E_0 = \hbar \omega_0 = \frac {\hbar^2} {M x_0^2} = \frac {\hbar^2} {M a a_{\uparrow \downarrow}} 
  \frac {4 e^{-2 \gamma - 1}} {\ln(a_{\uparrow \downarrow}/a)}.
\end{equation}
Finally, the number of atoms is measured in units of $N_0$, where
\begin{eqnarray}
  N_0 = \Psi_0^2 x_0^2 = \frac 1 {4 \pi} \ln^2(a_{\uparrow \downarrow}/a).
\label{n0}
\end{eqnarray}
In the rest of the manuscript we work in dimensionless units (using the units presented
above), while we give some estimates for the experimentally-relevant quantities in Sec.\,V.

We choose to work with fixed $L$ and $N$, minimizing the following extended energy functional 
\cite{GO}, which (in dimensionless units) takes the form \cite{PA}  
\begin{eqnarray}
  {\cal E}(\Psi, \Psi^*) = 
   \int \left( \frac {1} {2} |\nabla \Psi|^2 
  + \frac 1 2 \omega^2 \rho^2 (1 + \lambda \frac {\rho^2} {a_0^2}) |\Psi|^2 \right.
  \nonumber \\
  \left. + \frac 1 2 |\Psi|^4 \ln \frac {|\Psi|^2} {\sqrt{e}} \right) \, d^2 \rho
\nonumber \\
  - \mu \int \Psi^* \Psi \, d^2 \rho - \Omega \int \Psi^* {\hat L} \Psi \, d^2 \rho.
\label{funncc}
\end{eqnarray}
In the above equation $\Psi$ is normalized to the total number of atoms, $\int |\Psi|^2 
\, d^2 r = N$. Also, ${\hat L}$ is the operator of the angular momentum, while $\mu$ 
and $\Omega$ are Lagrange multipliers, corresponding to the conservation of the atom 
number and of the angular momentum, respectively. The corresponding nonlinear equation that $\Psi(\rho, \theta)$ satisfies is
\begin{equation}
 \left( - \tfrac 1 2 \nabla^2 + \tfrac 1 2 \omega^2 \rho^2 (1 + \lambda \tfrac {\rho^2} {a_0^2})
 + |\Psi|^2 \ln |\Psi|^2 - \Omega {\hat L} \right) \Psi = 
  \mu \Psi. 
 \label{nlin}  
\end{equation}

Equation (\ref{nlin}) was solved by minimizing numerically the functional of Eq.\,(\ref{funncc}), using the damped, second-order in fictitious time, method \cite{GO}, which is a method for constrained minimization. In our calculations, we used a square spatial grid, with $\delta x = \delta y = 0.1$, which was proven to be accurate enough, in the sense that it produced results that are converged with respect to the grid resolution. The size of the calculational domain was larger than presented in the figures below, in order to avoid boundary effects. 

We used a variety of trial order parameters as the initial condition for the calculations, namely, states that represent center-of-mass excitation, surface-wave excitation and vortex excitation, as well as ``mixed" states, which correspond to combinations of the aforementioned modes of excitation. The use of multiple initial conditions in the calculation, for each value of the angular momentum, and the comparison of the corresponding energies of the solutions, was necessary to verify that we reached the lowest-energy state, and not some local minimum of the energy functional, which would correspond to an excited state. 

\section{Results}

Given that there are many parameters, in the derived results we consider the 
case where both the harmonic, as well as the anharmonic terms in the energy are 
smaller than the energy that results from the nonlinear term. In the opposite limit 
the droplet is ``squeezed" by the trap and the physics is -- at least qualitatively 
-- similar to the one-component system, with an effective repulsive interaction.

We have performed extensive numerical simulations and below we present some
representative data, for four values of $N = 50$, $100$, $150$ and $200$, for a fixed
value of $\lambda = 0.05$, while $\omega$ is also fixed and equal to 0.05.
For a free droplet in the Thomas-Fermi limit, the radial 
size of the droplet $\rho_0$ is $\sqrt{N {\sqrt e}/\pi}$. For $N \approx 100$,
which is the typical $N$ that we use, $\rho_0 \approx 10$. On the other hand,
the oscillator length $a_0 = 1/\sqrt{\omega}$ is $\approx 5$, i.e., the two 
length scales are comparable (as they should be). Finally, the anharmonic term 
in the energy $\lambda \rho^2/a_0^2$ is on the order $\lambda \omega N$, which 
is somewhat less than unity.

\subsection{N = 50}

We start with a ``small" (scaled) atom number, $N = 50$. In a purely harmonic potential
the center-of-mass coordinate separates from the relative coordinates and these
two degrees of freedom are decoupled. As a result, one way for the droplet to 
carry its angular momentum is via center-of-mass excitation. This is actually 
what happens for $N = 50$ and $\lambda = 0$ \cite{NKO}. 

\begin{figure}
\includegraphics[width=\columnwidth ,angle=-0]{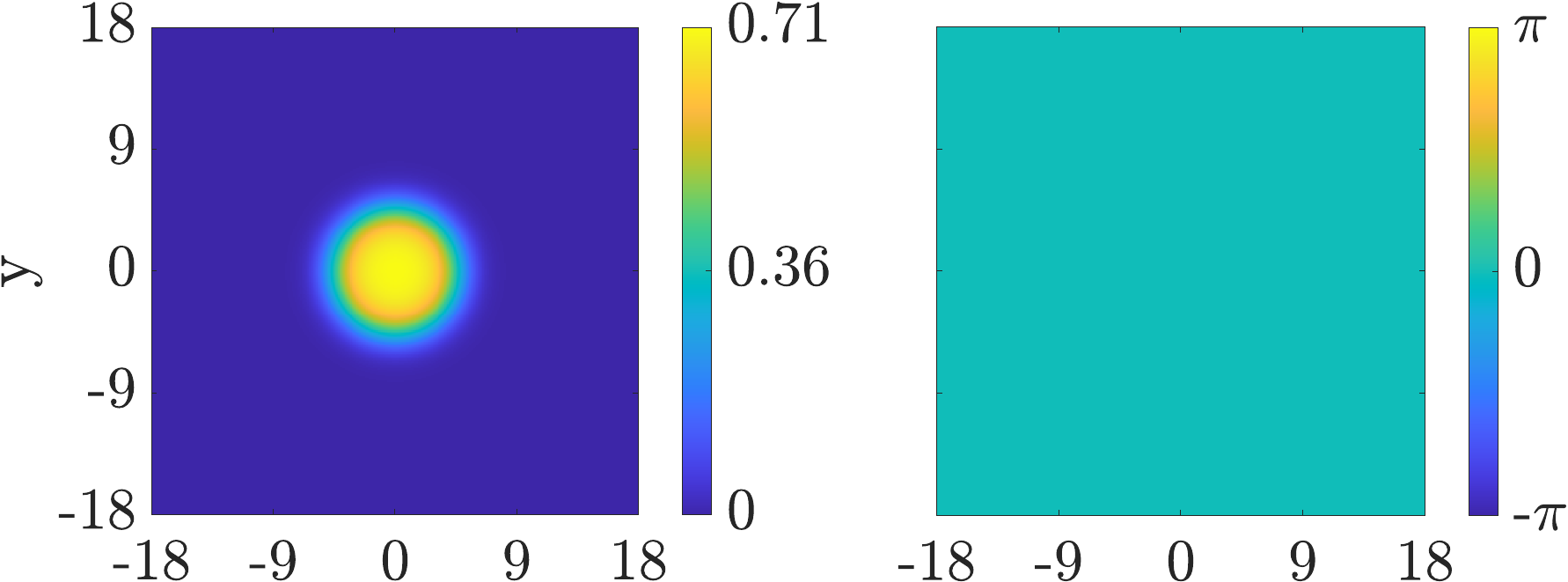}\\
\vspace{0.3\baselineskip}
\includegraphics[width=\columnwidth ,angle=-0]{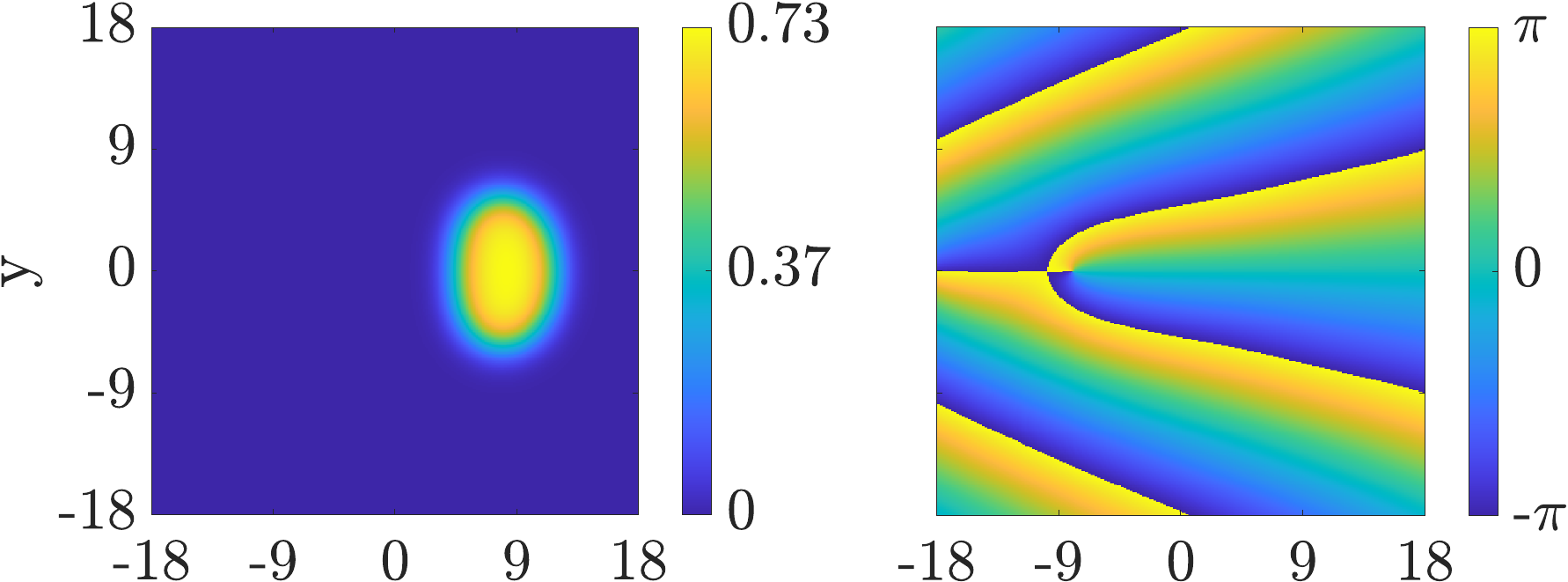}\\
\vspace{0.3\baselineskip}
\includegraphics[width=\columnwidth ,angle=-0]{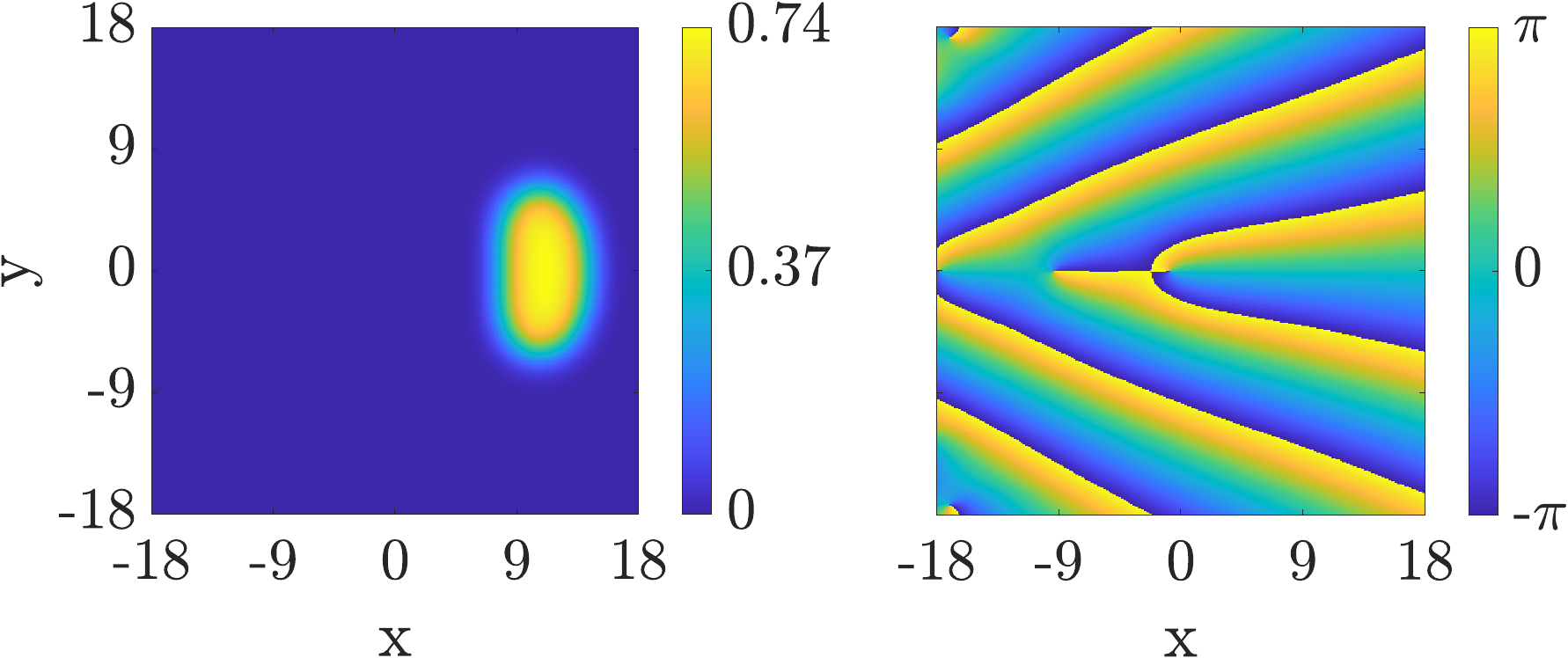}\\
\vspace{\baselineskip}
\includegraphics[width=\columnwidth ,angle=-0]{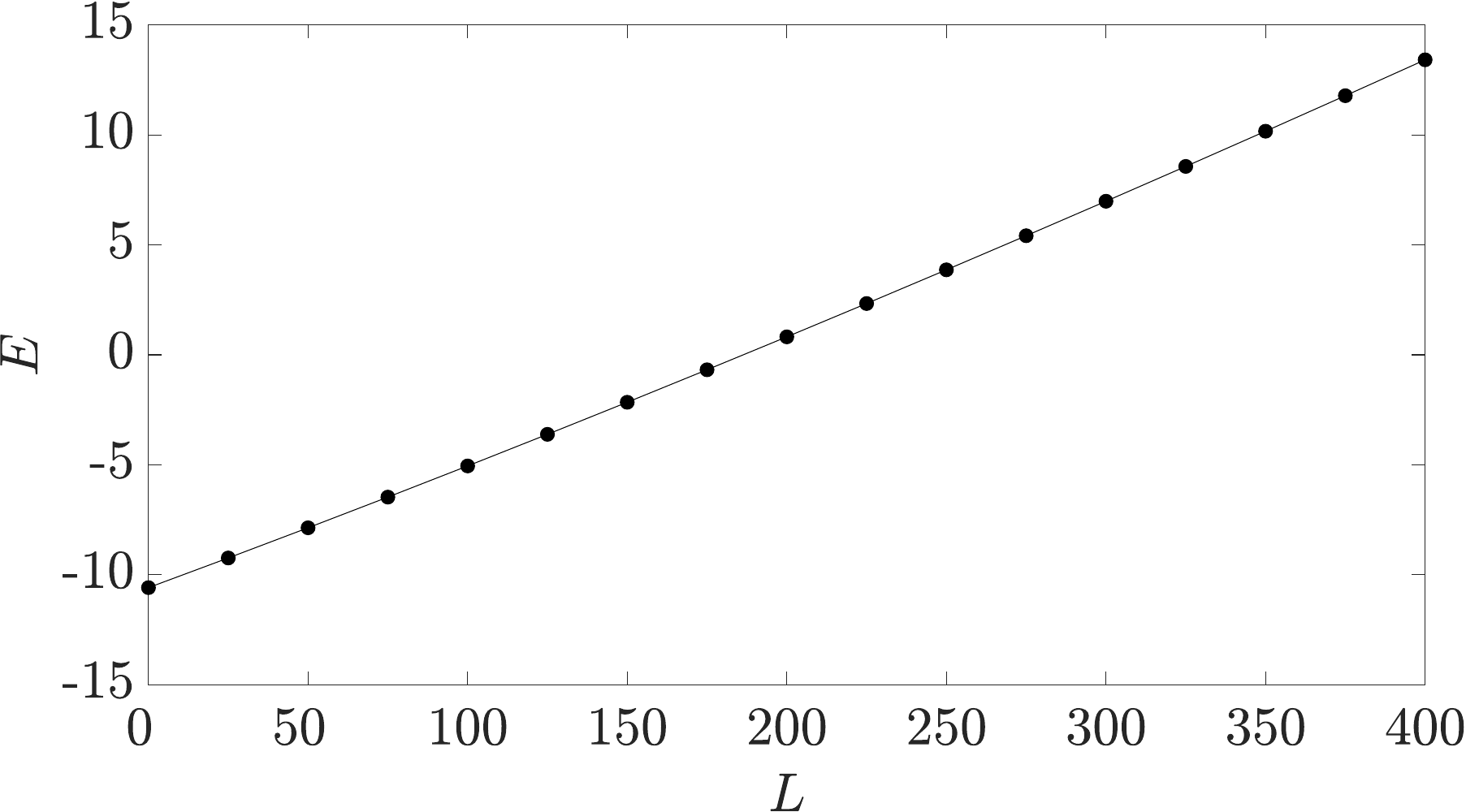}
\caption{Upper plots: The density (left column) and the phase 
(right column) of the droplet order parameter, for $N = 50$, $\omega = 0.05$, 
$\lambda = 0.05$, and $\ell = 0.0$, $4.0$, and 8.0, from top to bottom. Here 
the density is measured in units of $\Psi_0^2$ and the length in units of $x_0$. 
Lower plot: The corresponding dispersion relation as function of $L$. Here 
the energy is measured in units of $E_0$ and the angular momentum in units of 
$\hbar$.}
\end{figure}

In the anharmonic potential that we have considered here the two kinds of 
excitation are coupled. As seen in Fig.\,1, we still have a picture that 
resembles center-of-mass excitation, however the droplet is also distorted 
from being exactly axially-symmetric. This is due to the presence of the quartic 
term in the trapping potential, which implies that the separation between 
center-of-mass and relative coordinates is no longer an exact result. 

\begin{figure}
\includegraphics[width=\columnwidth ,angle=-0]{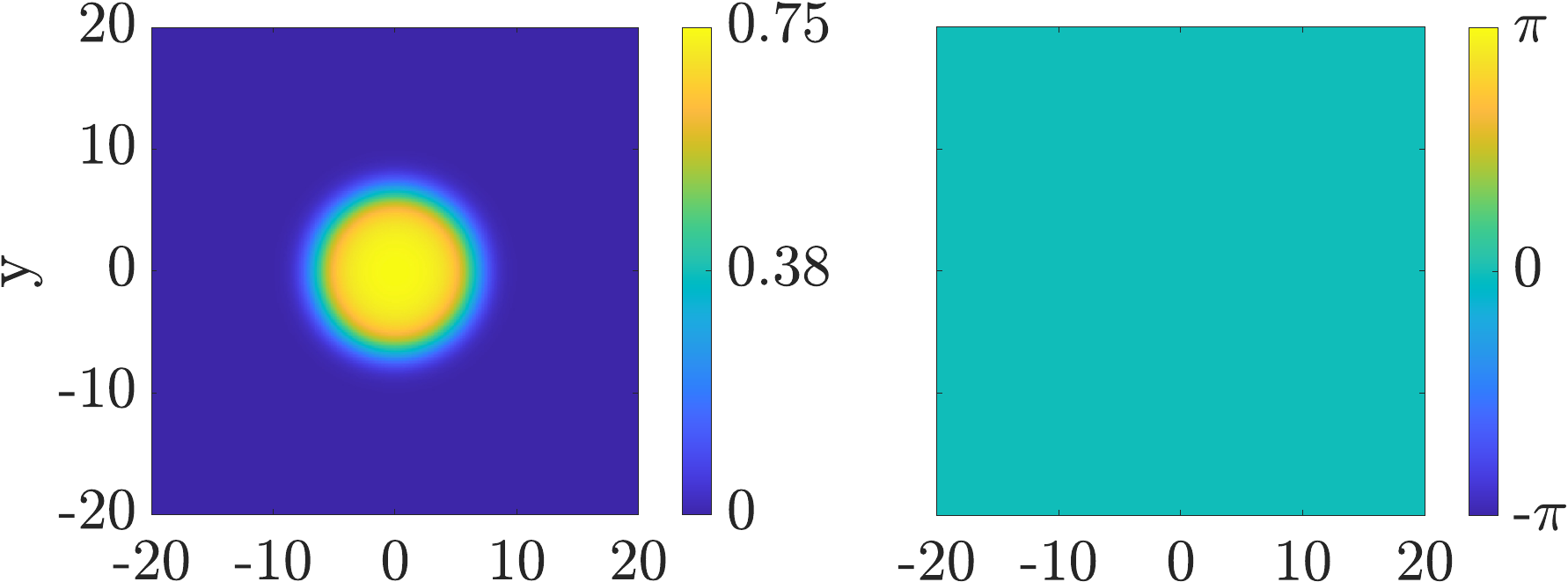}\\
\vspace{0.3\baselineskip}
\includegraphics[width=\columnwidth ,angle=-0]{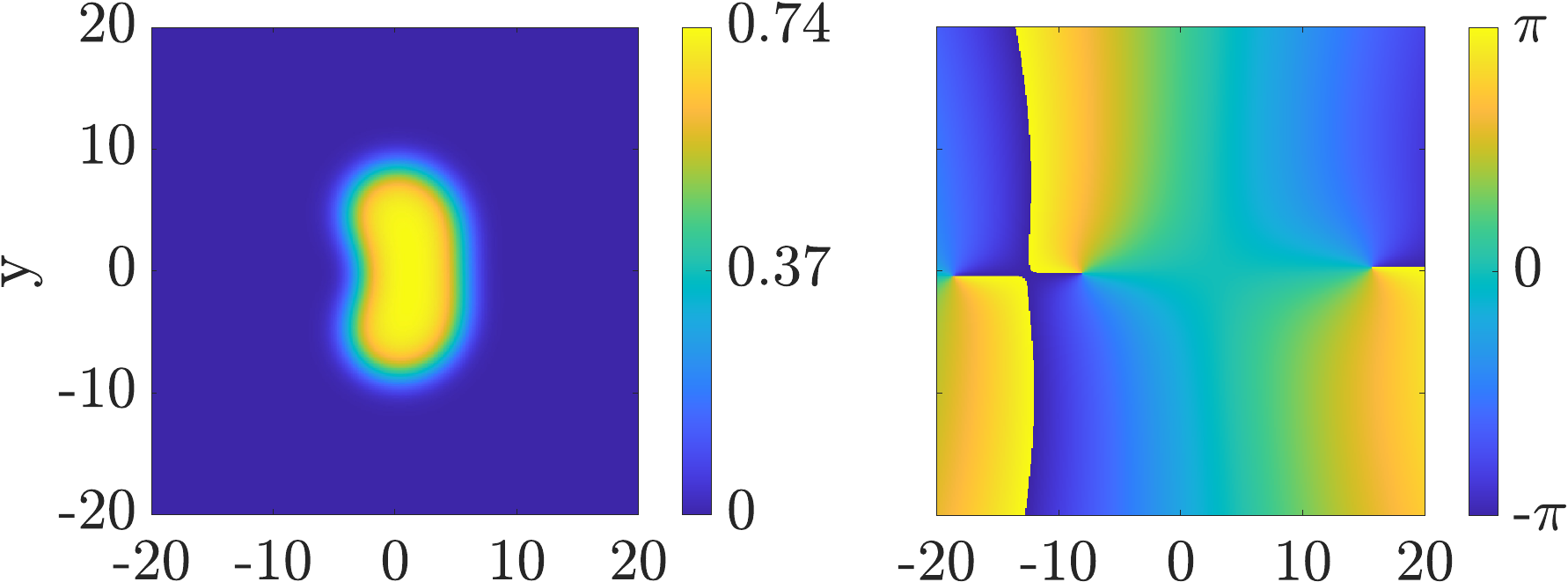}\\
\vspace{0.3\baselineskip}
\includegraphics[width=\columnwidth ,angle=-0]{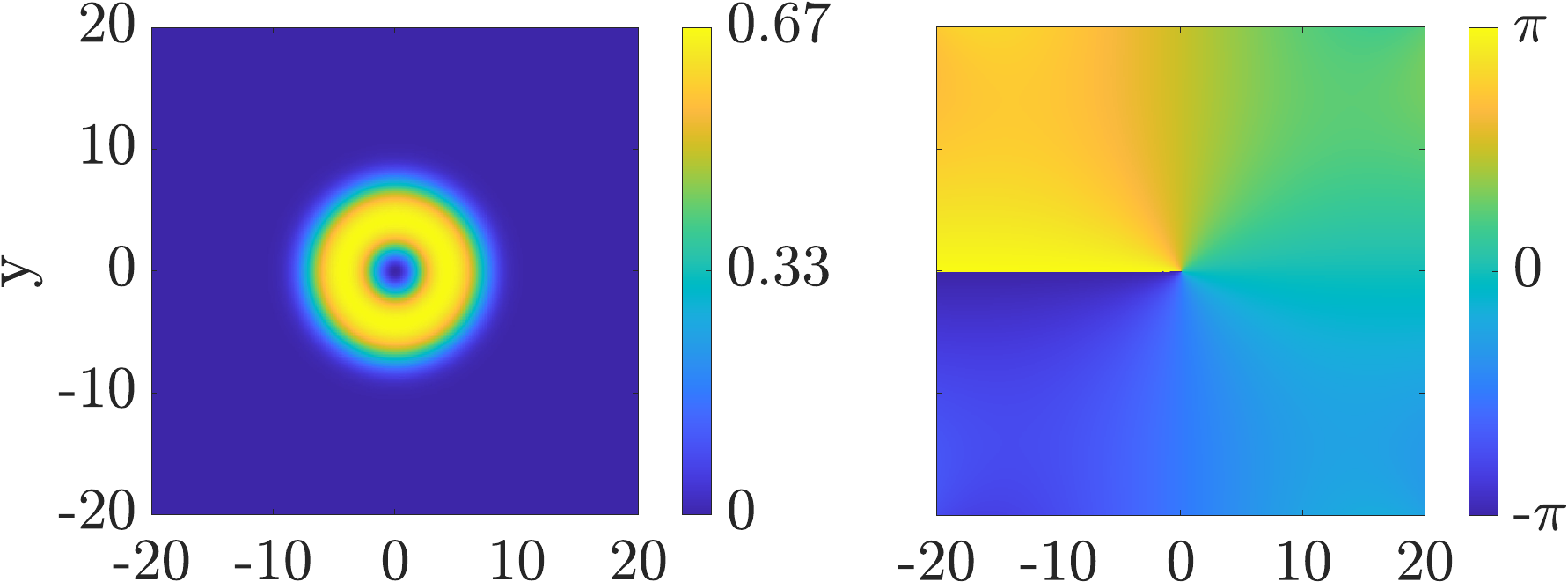}\\
\vspace{0.3\baselineskip}
\includegraphics[width=\columnwidth ,angle=-0]{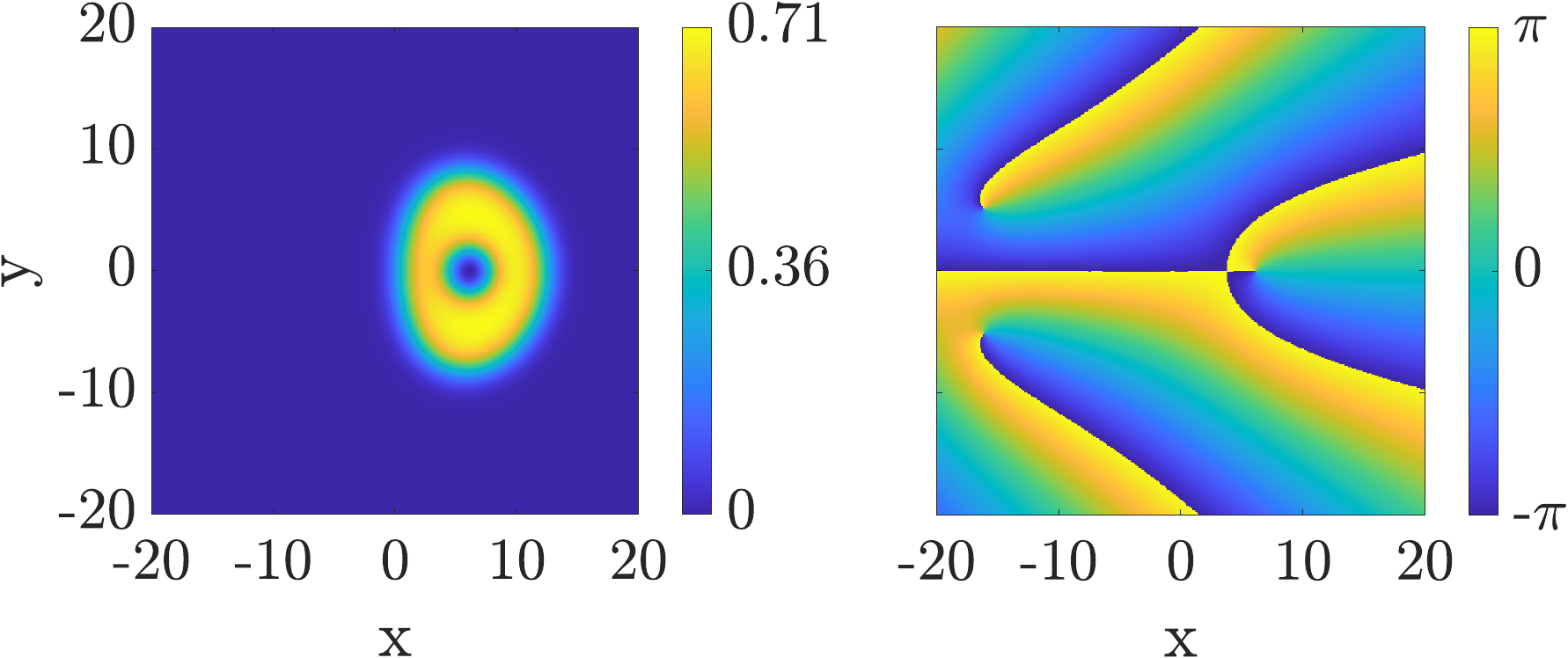}\\
\caption{The density (left column) and the phase 
(right column) of the droplet order parameter, for $N = 100$, $\omega = 0.05$, 
$\lambda = 0.05$, and $\ell = 0.0$, $0.5$, $1.0$, and $3.5$, from
top to bottom. Here the density is measured in units of $\Psi_0^2$ and the length 
in units of $x_0$.}
\addtocounter{figure}{-1}
\end{figure}
\begin{figure}
\includegraphics[width=\columnwidth ,angle=-0]{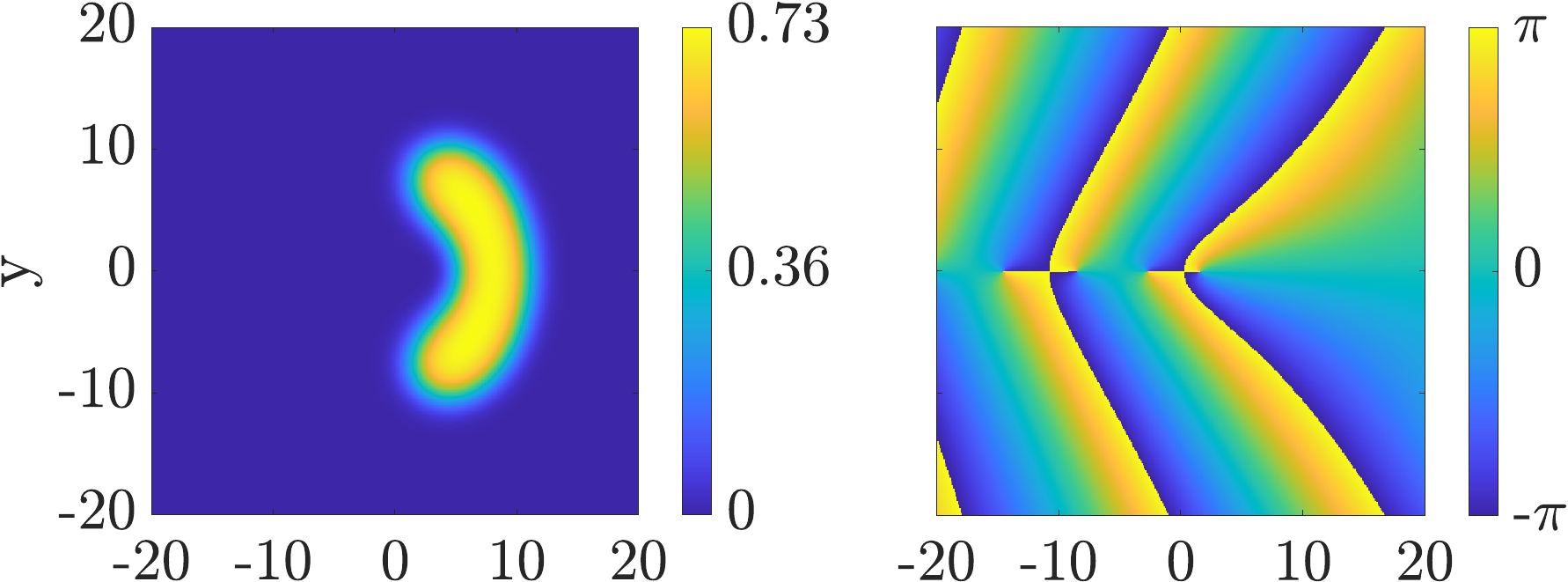}\\
\vspace{0.3\baselineskip}
\includegraphics[width=\columnwidth ,angle=-0]{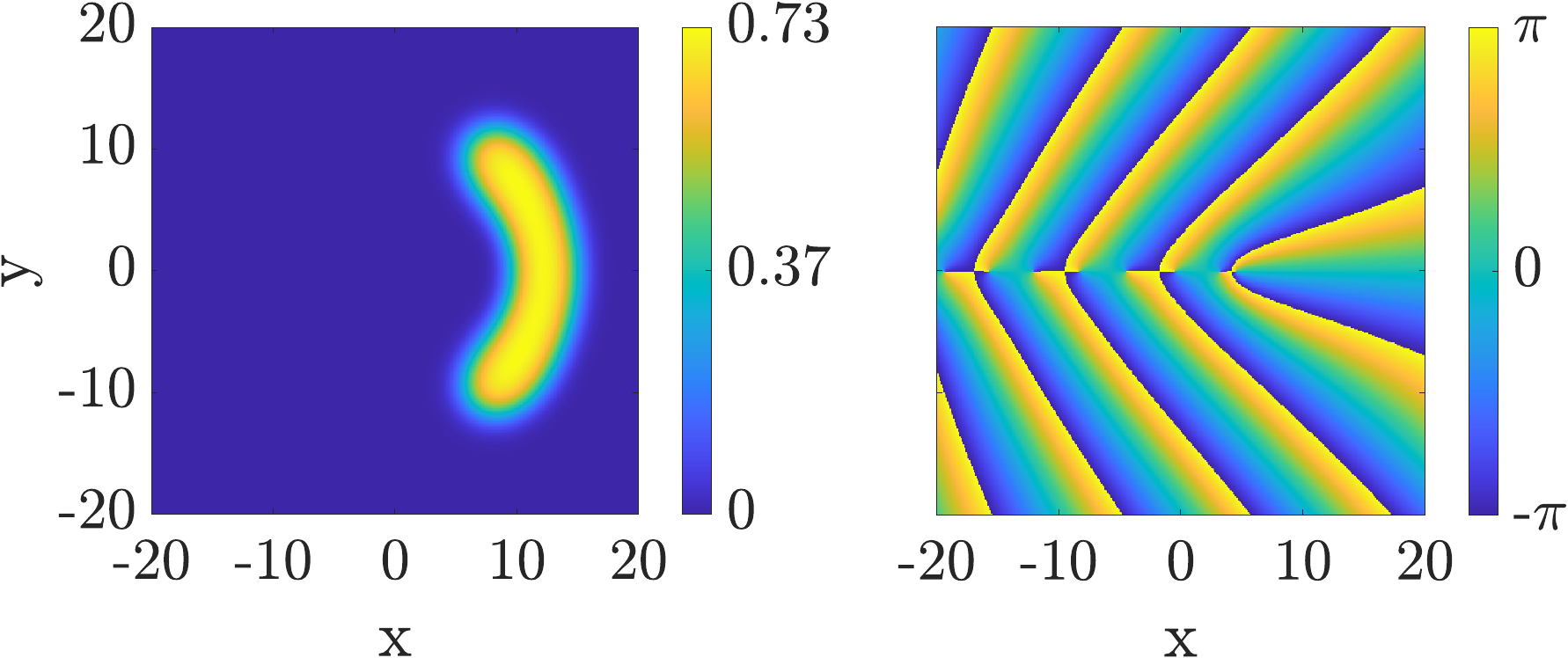}\\
\vspace{\baselineskip}
\includegraphics[width=\columnwidth ,angle=-0]{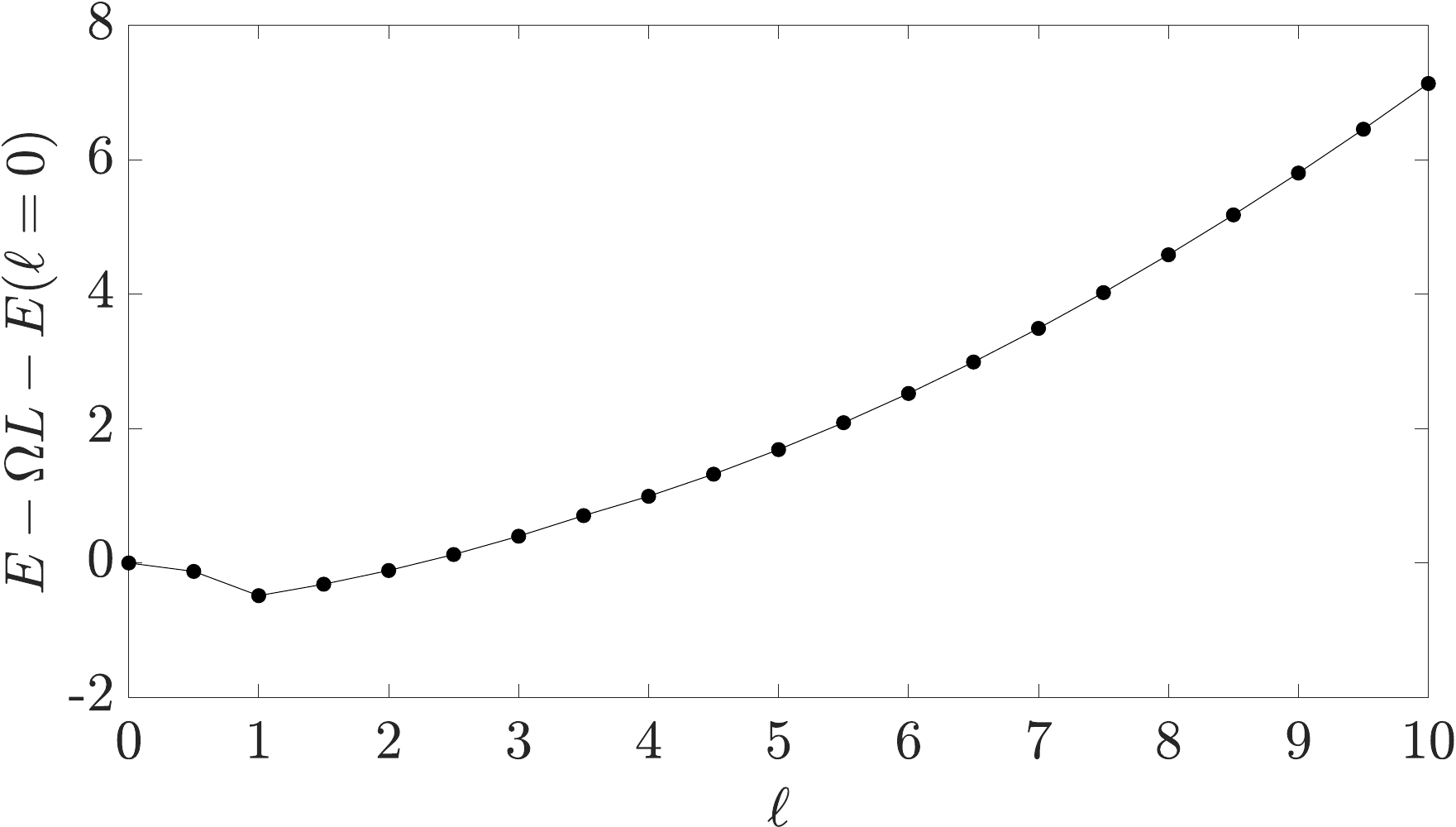}
\caption{(Cont.) Upper plots: The density (left column) and the phase 
(right column) of the droplet order parameter, for $N = 100$, $\omega = 0.05$, 
$\lambda = 0.05$, and $\ell = 4.0$ (top), and $10.0$ (bottom). Here the density is measured in units of $\Psi_0^2$ and the length 
in units of $x_0$. Lower plot: The corresponding dispersion relation, in the 
rotating frame, i.e., $E_{\rm rot}(\ell) - E(\ell = 0)$ as function of $\ell$, with 
$\Omega = 0.054$. Here the energy is measured in units of $E_0$ and the 
angular momentum in units of $\hbar$.}
\end{figure}

Another effect of the quartic term in the confining potential is that the 
effective potential, i.e., the trapping plus the centrifugal, 
\begin{eqnarray}
 V_{\rm eff} = V(\rho) - \frac 1 2 \Omega^2 \rho^2 = 
 \frac 1 2 (\omega^2 - \Omega^2) \rho^2 + \frac {\lambda} 2 \omega^2 \frac {\rho^4} {a_0^2}, 
\end{eqnarray}
takes the form of a ``Mexican-hat" for $\Omega > \omega$. Its minimum occurs at 
\begin{eqnarray}
  \frac {\rho_0} {a_0} = \left[ \frac 1 {2 \lambda} \left( \frac {\Omega^2} {\omega^2} - 1 \right) \right]^{1/2}.
\end{eqnarray} 
For the data shown in Fig.\,1, when $\ell = 4.0$, then $\Omega \approx 0.06024$, and 
the above equation gives $\rho_0 \approx 9.5$, while for $\ell = 8.0$, where 
$\Omega \approx 0.06556$, $\rho_0 \approx 12.0$. These values of $\rho_0$ coincide with
the minimum of the effective potential and this is what determines the location 
of the droplet. As mentioned also earlier, the presence of a quartic term in the 
potential plays a crucial role. If this is not present, for $\Omega > \omega$ there 
is no restoring force and the droplet would escape to infinity.

Let us turn to the dispersion relation $E(L)$, which we show at the bottom plot 
of Fig.\,1. Fitting the numerical data with a quadratic polynomial we find that 
\begin{equation}
 E(L) \approx -10.6032 + 0.054117 L + 1.4911 \times 10^{-5} L^2.
 \label{quadfit}
\end{equation}
We see that $E(L)$ is almost a linear function, with a slope which is higher than 
$\omega$, though. Also, the curvature is small and positive. These results are 
analysed below.

\subsection{N = 100}

The second value of $N$ that we consider is 100. As seen in Fig.\,2, in this case 
the droplet carries its angular momentum in a very different way. For values 
of the angular momentum $0 < \ell < 1$, the droplet gets distorted due to the 
approach of a vortex state. For $\ell = 1$ there is a singly-quantized vortex state 
that is located at the center of the trap, and of the droplet. For higher values of 
$\ell$, the droplet starts to move away from the center of the trap, in a ``mixed" 
state, which resembles center-of-mass excitation of the vortex-carrying droplet. However, this ``mixed" state again has a density distribution that is axially-asymmetric. Specifically, the inner half of the droplet (i.e., the one closer to the origin) gets progressively more ``squeezed", as the value of $\ell$ increases.

As $\ell$ increases even further, the situation changes completely. For 
$\ell = 4.0$, it is no longer energetically favourable for the droplet to accommodate 
a vortex in the distorted ``mixed" state. Rather, it takes advantage of the ``Mexican-hat" shape of the 
effective potential, which has a minimum at $\rho_0 \approx 9.5$ (for $\Omega \approx 
0.060131$), while $\rho_0$ becomes $\approx 13.0$ for $\ell = 10.0$ (where $\Omega \approx 
0.067842$). 

The corresponding energy, which is shown at the bottom of Fig.\,2 in the rotating
frame (i.e., $E_{\rm rot} = E(\ell) - L \Omega$) for $\Omega = 0.054$, develops some structure in this case, contrary to Fig.\,1. More specifically, 
we see that there is a minimum for $\ell = 1$.

\subsection{N = 150}

The third value of $N$ that we consider is $150$. Here, when $0 < \ell \leq 1$, the picture is qualitatively the same as for $N = 100$. However, for $1 < \ell < 2$, a second vortex approaches, forming a doubly-quantized vortex state for $\ell = 2$. For larger $\ell$ values, in particular $2 < \ell \leq 4$, the droplet exists in a ``mixed" state, which now resembles center-of-mass-like excitation containing two singly-quantized vortices. Finally, for $\ell \geq 4.5$, the droplet forms again a localized state with center-of-mass-like excitation.

The corresponding energy, which is shown at the bottom of Fig.\,3 in the rotating frame, for $\Omega = 0.054$, again develops some structure. More specifically, we see that there are two minima, for $\ell = 1$ and $2$, corresponding to different values of $\Omega$.

\subsection{N = 200}

The fourth, and final, value of $N$ that we consider is $200$. In this case, we observe in
Fig.\,4 that when $\ell$ is sufficiently small, i.e., up to $\ell = 2.5$, the 
droplet carries its angular momentum via vortex excitation of single quantization.
For $\ell = 3.0$, we have a triply-quantized vortex state, instead. For $\ell = 3.5$
there are three singly-quantized vortices, with an asymmetric density distribution. This state belongs to the class of ``mixed'' states, which combine vortex and center-of-mass-like excitations. However, here, this mode of excitation is energetically favourable only for a small range of angular momentum values. Specifically, for $\ell = 4.0$,
there is, again, a multiply-quantized vortex state with winding number equal to four.   
For $\ell = 4.5$ we have ``phantom" vortices, i.e., vortex states of single 
quantization at the regions of space where the density is very low, with a 
droplet density which is distorted. For $\ell = 5.0$ there is a multiply-quantized 
vortex state with winding number equal to five. Interestingly enough, for $\ell = 
5.5$, the droplet density breaks the axial symmetry, forming a localized blob. 
For $\ell = 6.0$ we have a multiply-quantized vortex state with winding number 
equal to 6. Finally, for values of $\ell \geq 6.5$, the droplet forms again 
a localized blob, along the minima of the effective potential.

\begin{figure}[ht!]
\includegraphics[width=\columnwidth ,angle=-0]{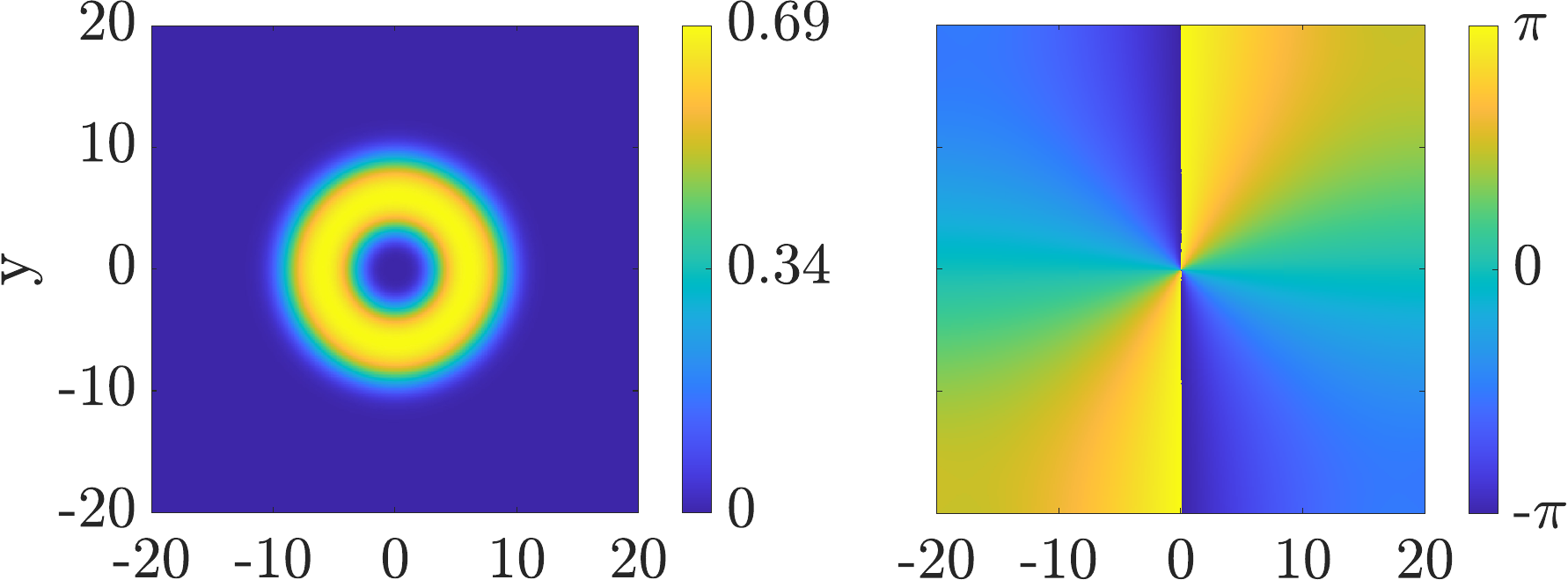}\\
\vspace{0.3\baselineskip}
\includegraphics[width=\columnwidth ,angle=-0]{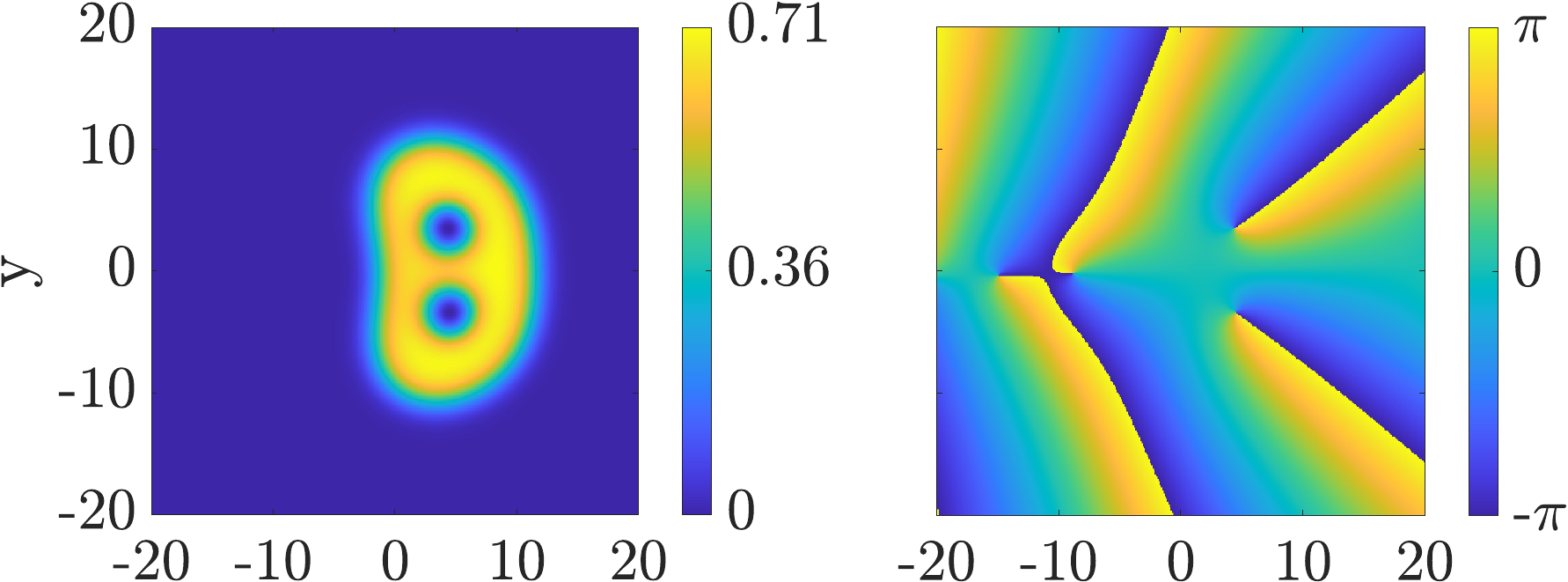}\\
\vspace{0.3\baselineskip}
\includegraphics[width=\columnwidth ,angle=-0]{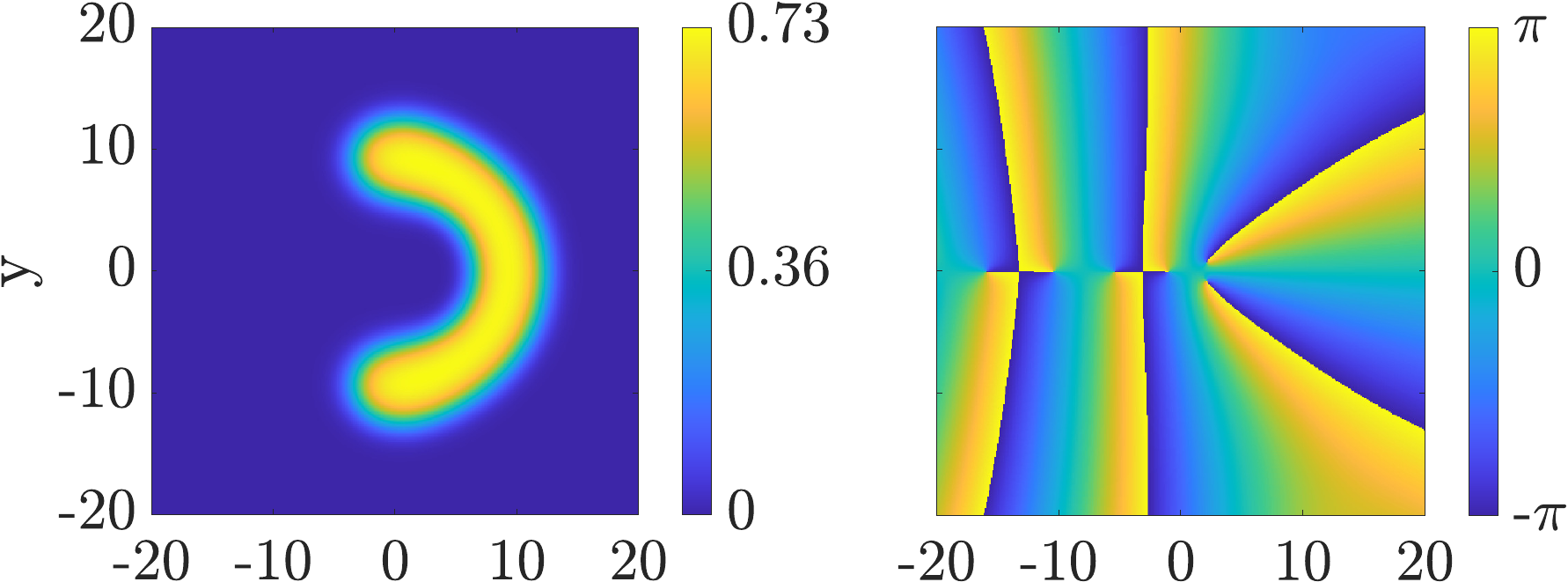}\\
\vspace{0.3\baselineskip}
\includegraphics[width=\columnwidth ,angle=-0]{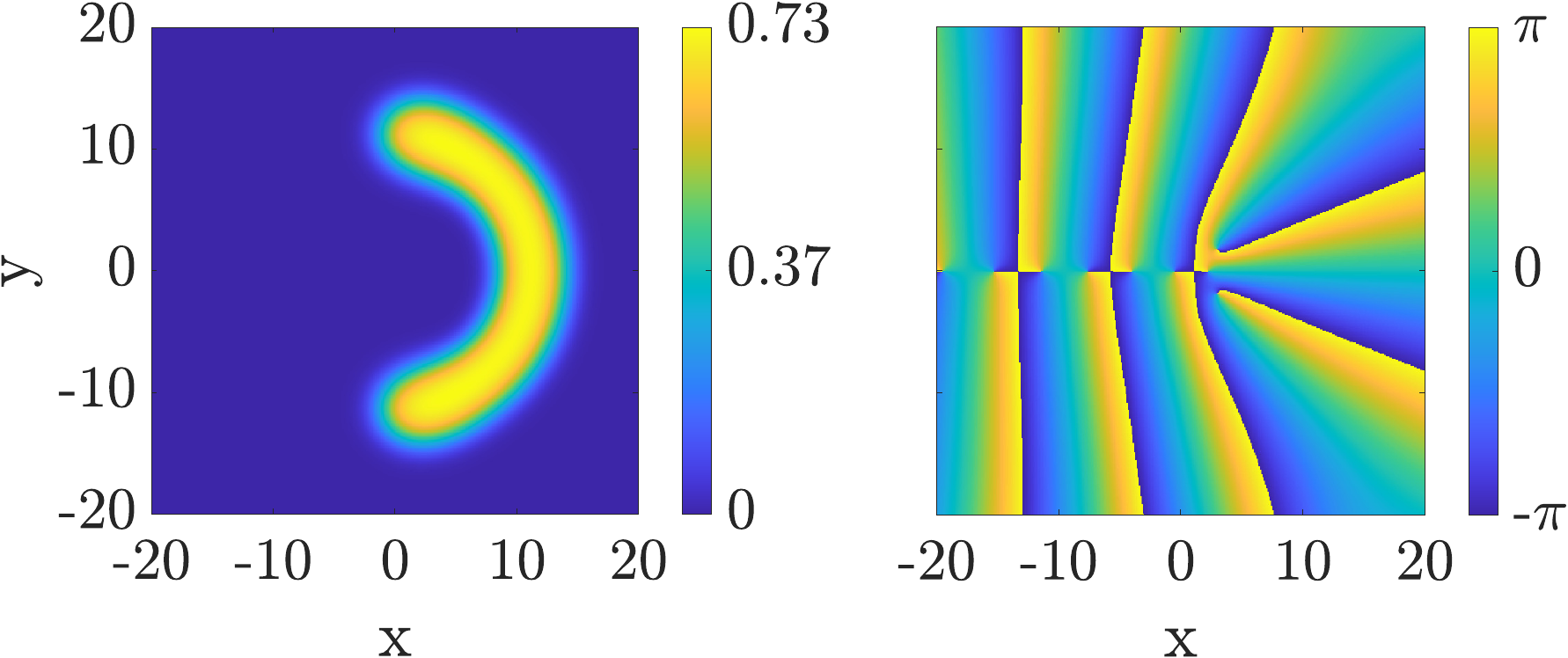}\\
\vspace{\baselineskip}
\includegraphics[width=\columnwidth ,angle=-0]{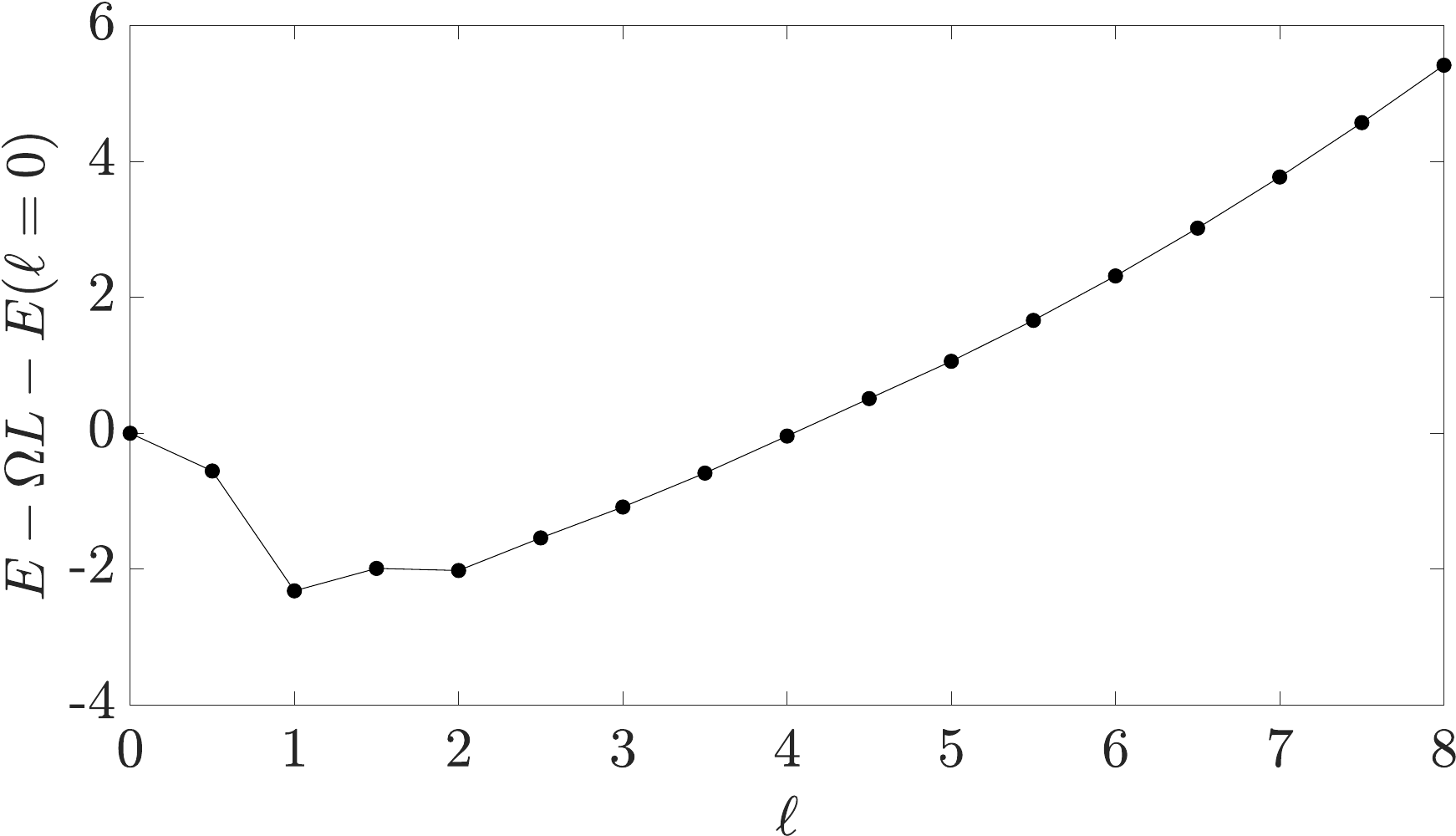}
\caption{Upper plots: The density (left column) and the phase 
(right column) of the droplet order parameter, for $N = 150$, $\omega = 0.05$, 
$\lambda = 0.05$, and $\ell = 2.0$, $3.5$, $5.0$, and $8.0$, from
top to bottom. Here the density is measured in units of $\Psi_0^2$ and the length 
in units of $x_0$. Lower plot: The corresponding dispersion relation, in the 
rotating frame, i.e., $E_{\rm rot}(\ell) - E(\ell = 0)$ as function of $\ell$, with 
$\Omega = 0.054$. Here the energy is measured in units of $E_0$ and the 
angular momentum in units of $\hbar$.}
\end{figure}

\begin{figure}
\includegraphics[width=\columnwidth ,angle=-0]{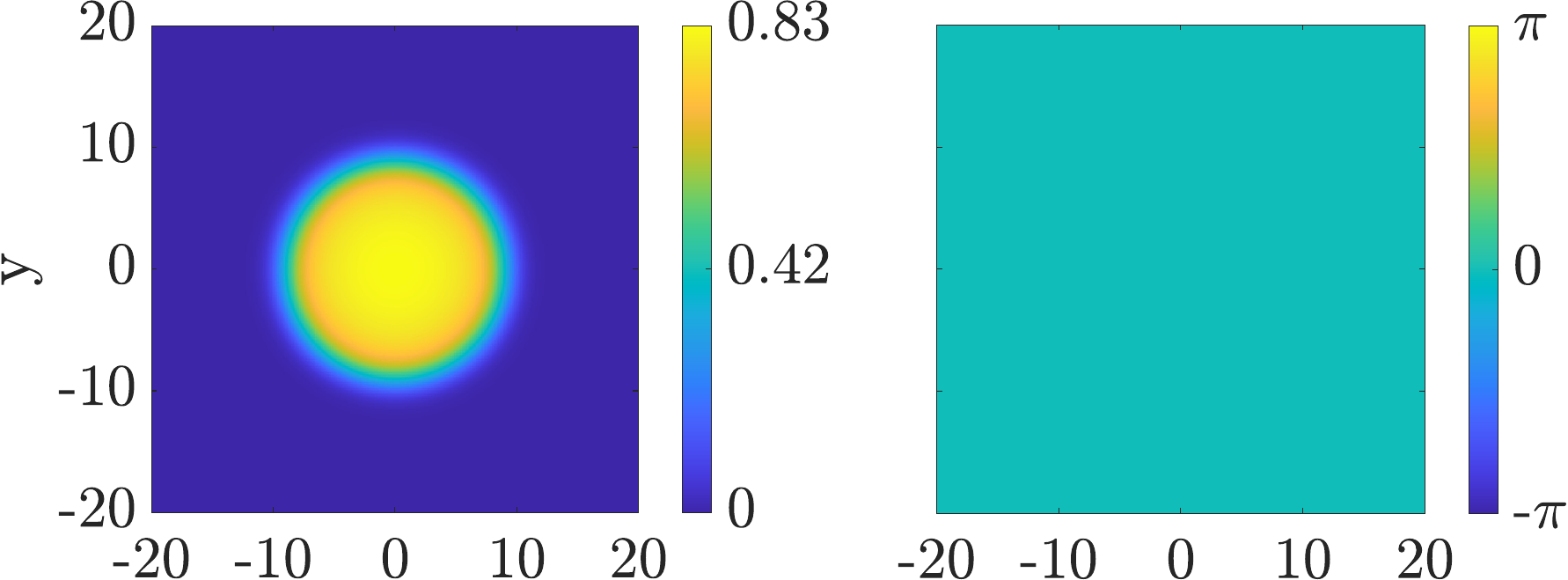}\\
\vspace{0.3\baselineskip}
\includegraphics[width=\columnwidth ,angle=-0]{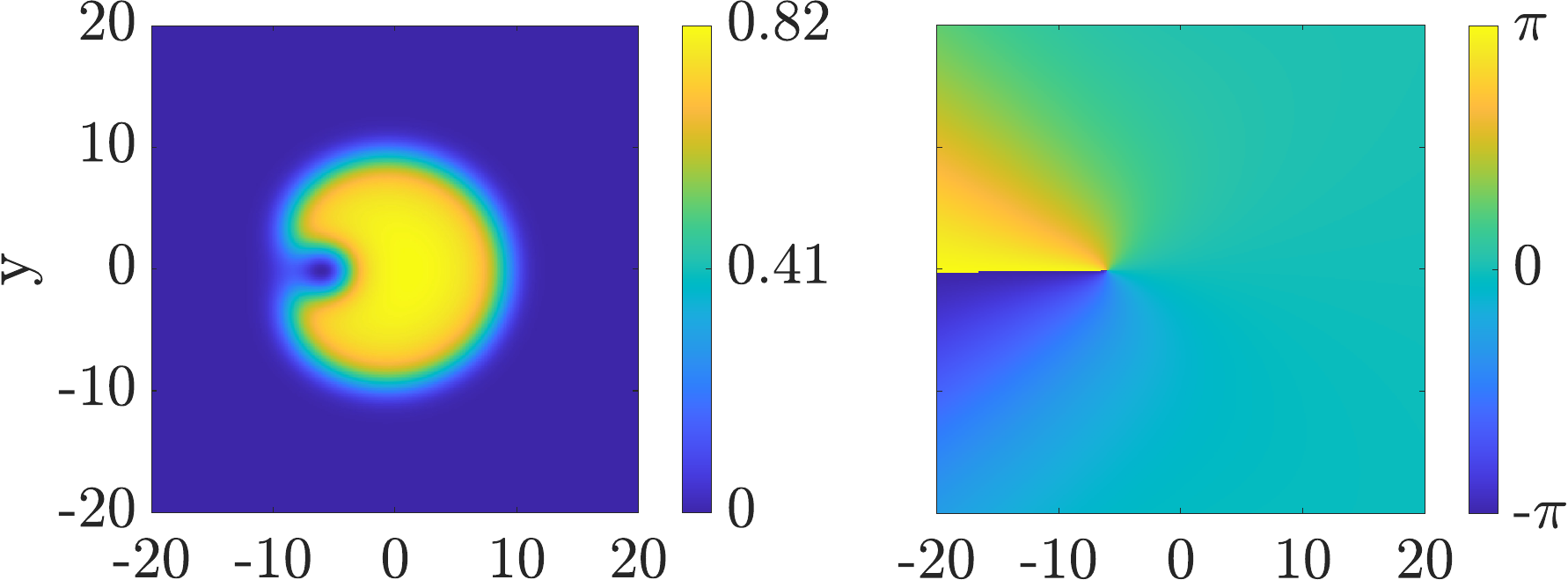}\\
\vspace{0.3\baselineskip}
\includegraphics[width=\columnwidth ,angle=-0]{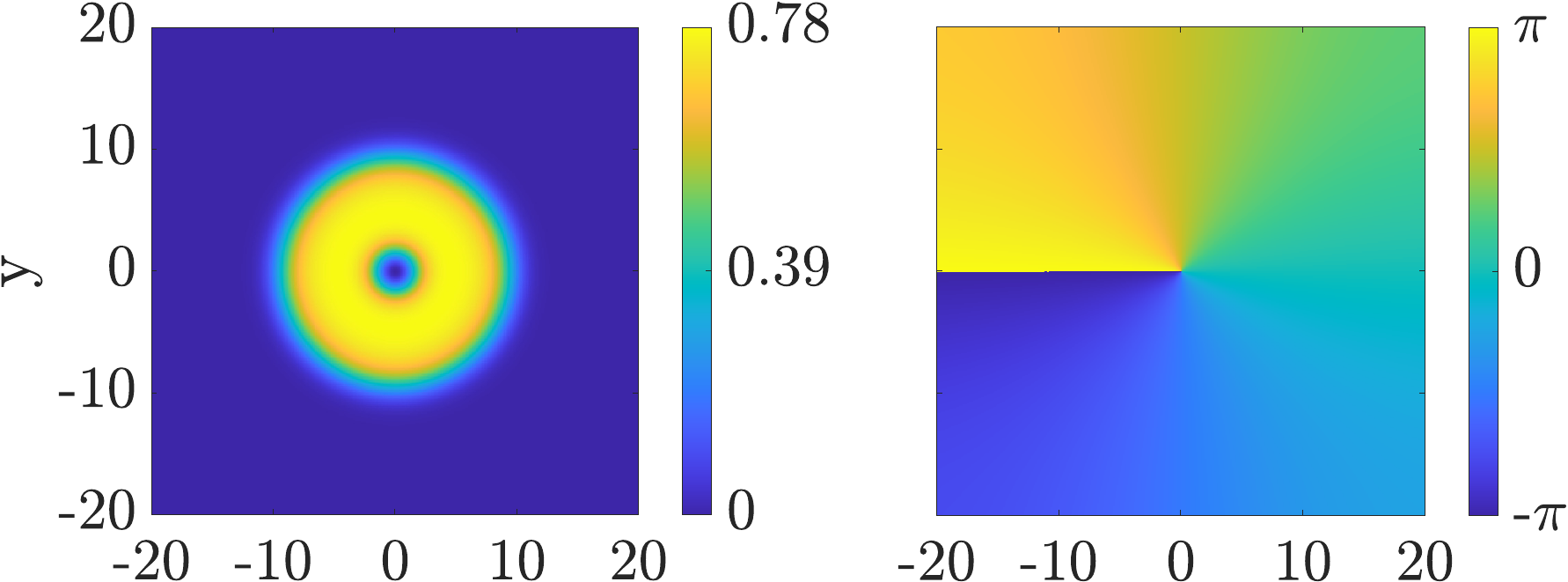}\\
\vspace{0.3\baselineskip}
\includegraphics[width=\columnwidth ,angle=-0]{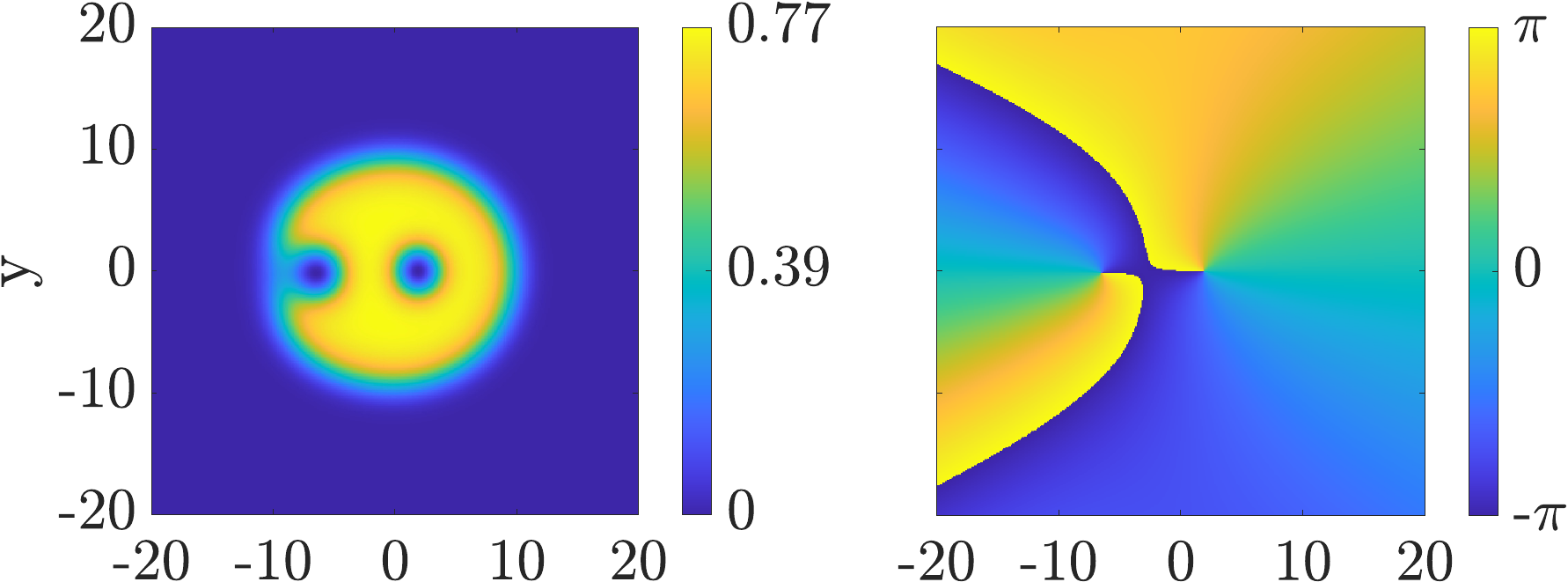}\\
\vspace{0.3\baselineskip}
\includegraphics[width=\columnwidth ,angle=-0]{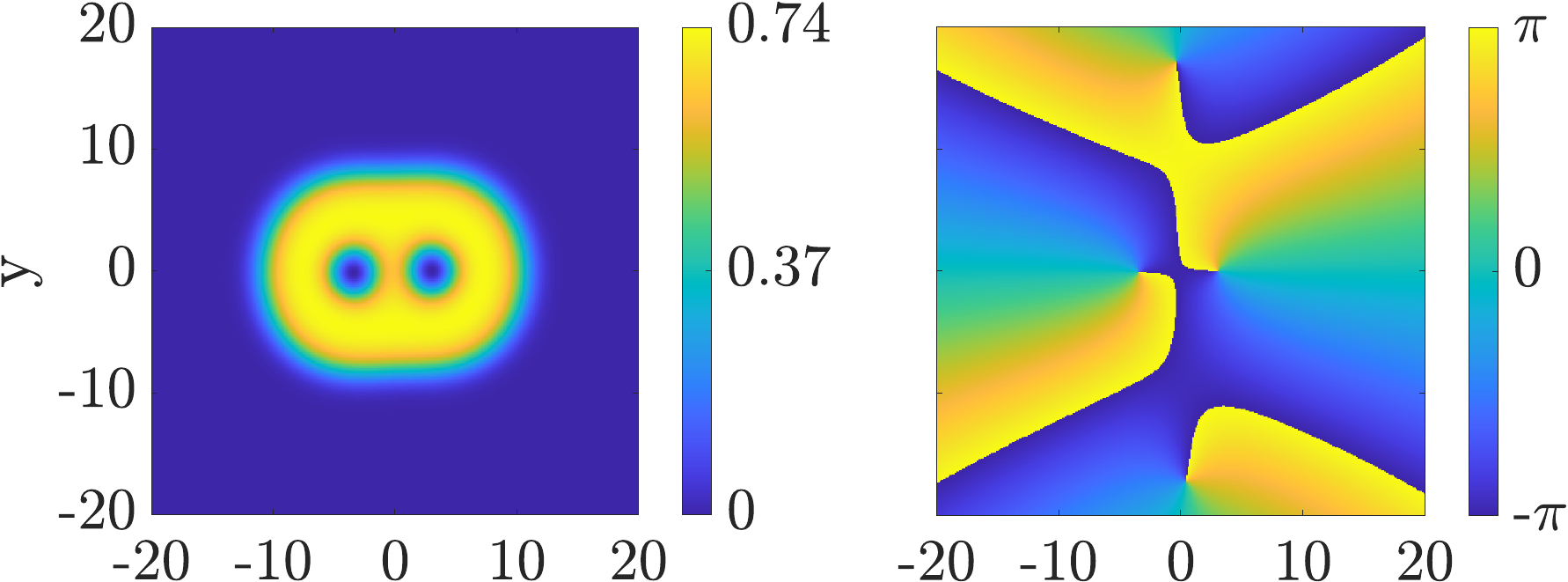}\\
\vspace{0.3\baselineskip}
\includegraphics[width=\columnwidth ,angle=-0]{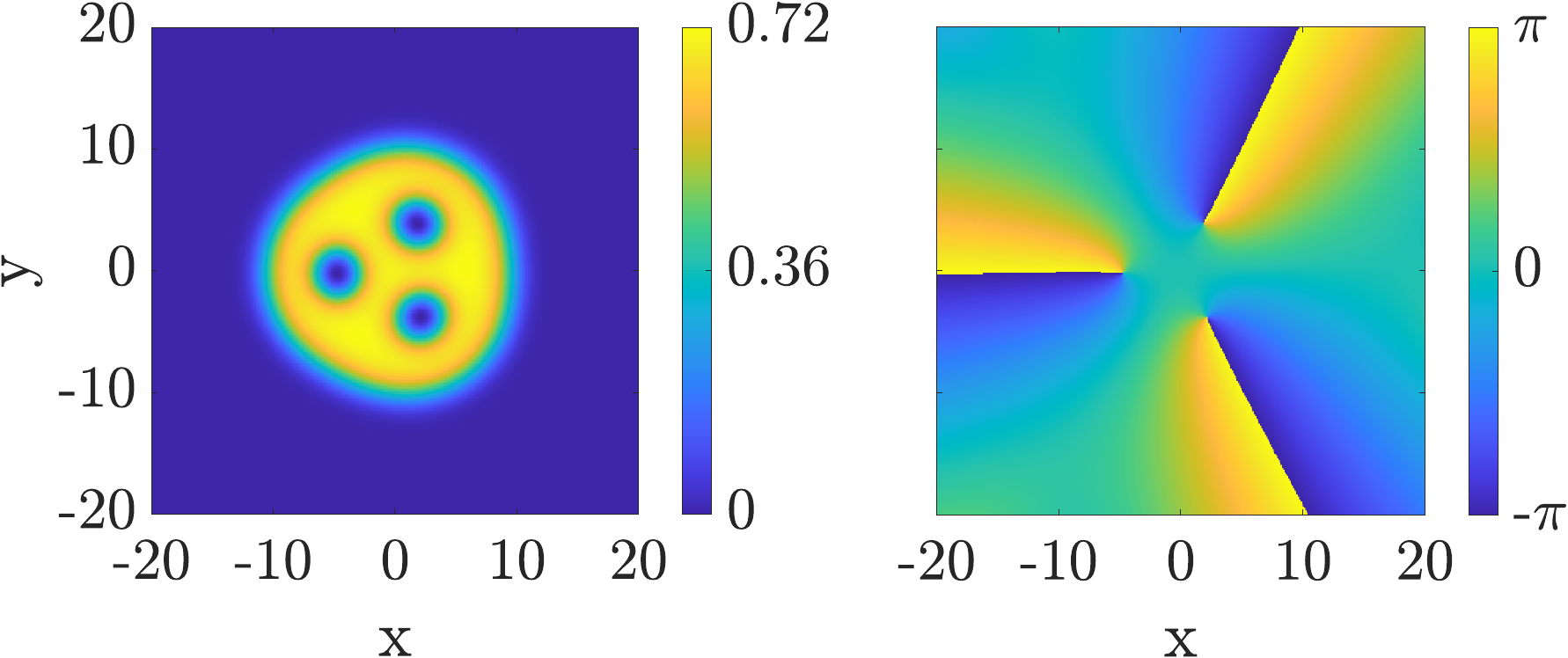}
\caption{The density (left column) and the phase 
(right column) of the droplet order parameter, for $N = 200$, $\omega = 0.05$, 
$\lambda = 0.05$, and $\ell = 0.0$, $0.5$, $1.0$, $1.5$, $2.0$, and $2.5$, from top to bottom. 
Here the density is measured in units of $\Psi_0^2$ and the length in units of 
$x_0$.}
\addtocounter{figure}{-1}
\end{figure}
\begin{figure}
\includegraphics[width=\columnwidth ,angle=-0]{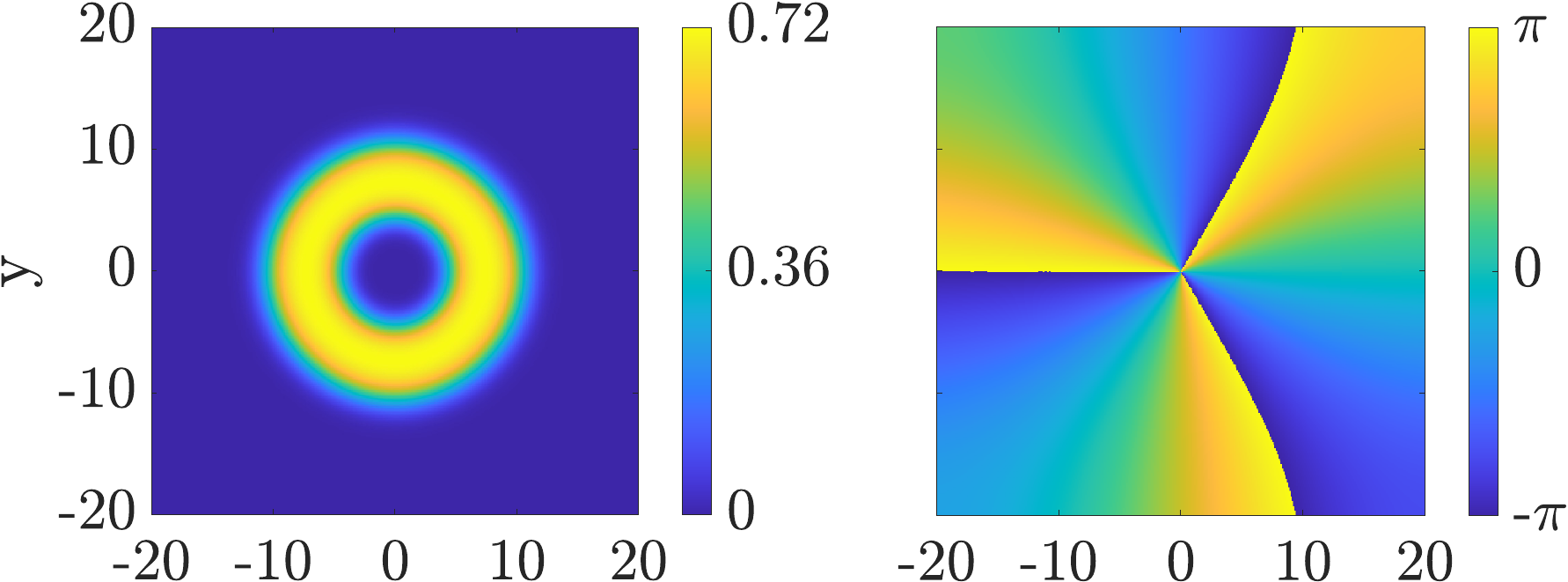}\\
\vspace{0.3\baselineskip}
\includegraphics[width=\columnwidth ,angle=-0]{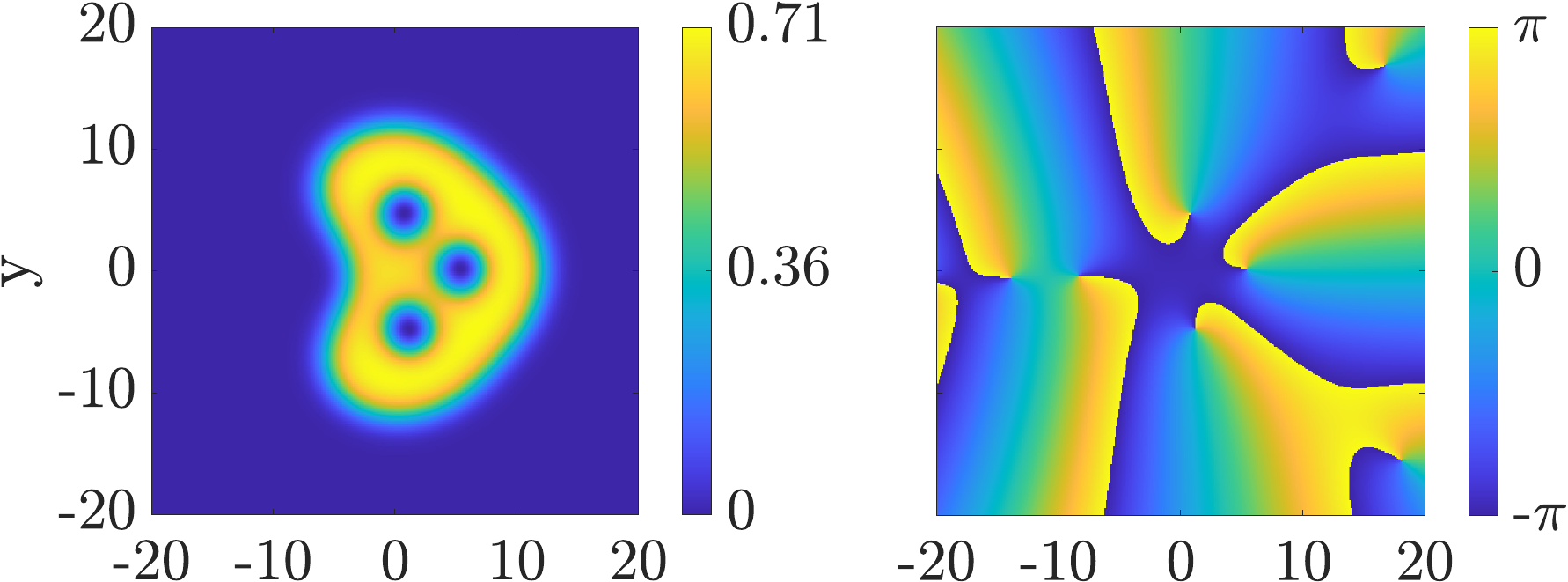}\\
\vspace{0.3\baselineskip}
\includegraphics[width=\columnwidth ,angle=-0]{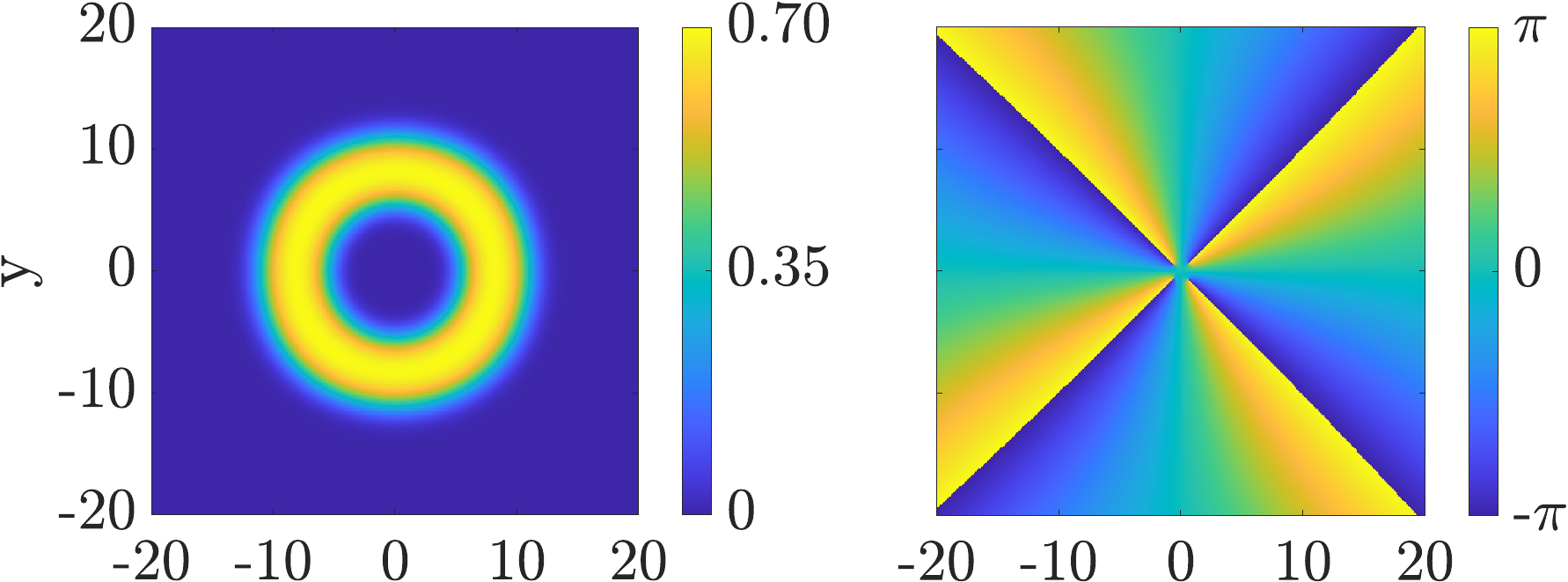}\\
\vspace{0.3\baselineskip}
\includegraphics[width=\columnwidth ,angle=-0]{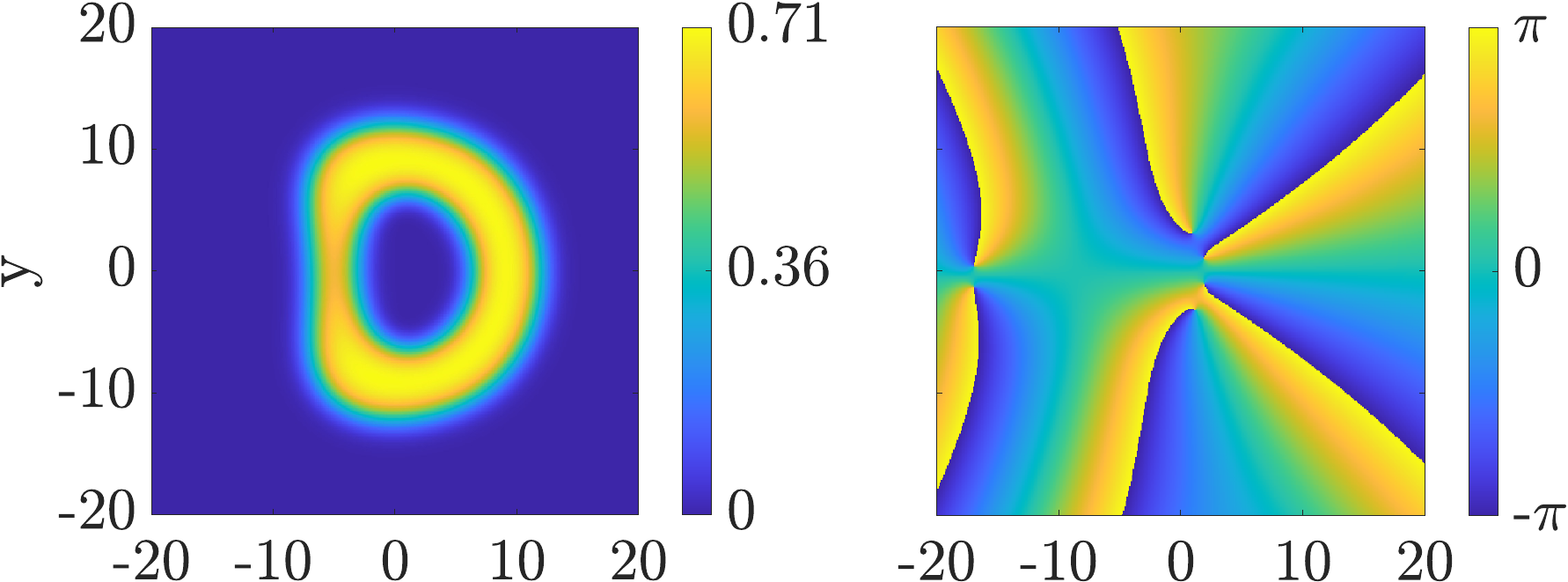}\\
\vspace{0.3\baselineskip}
\includegraphics[width=\columnwidth ,angle=-0]{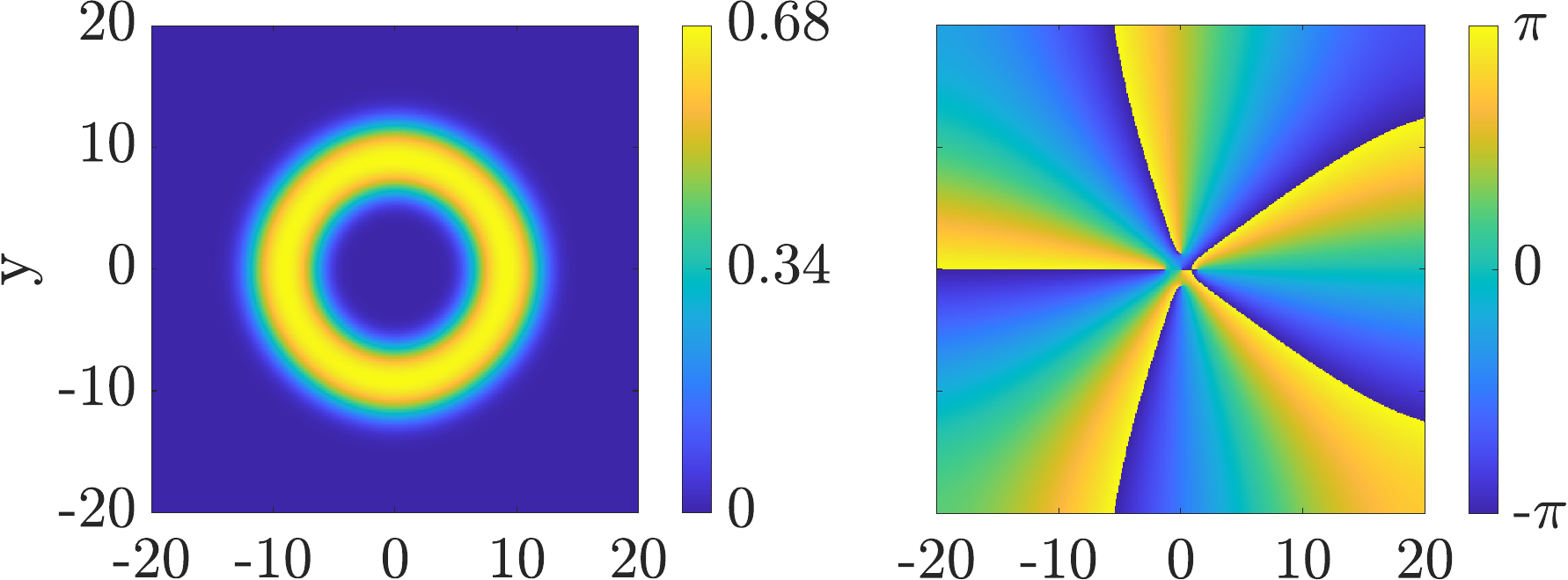}\\
\vspace{0.3\baselineskip}
\includegraphics[width=\columnwidth ,angle=-0]{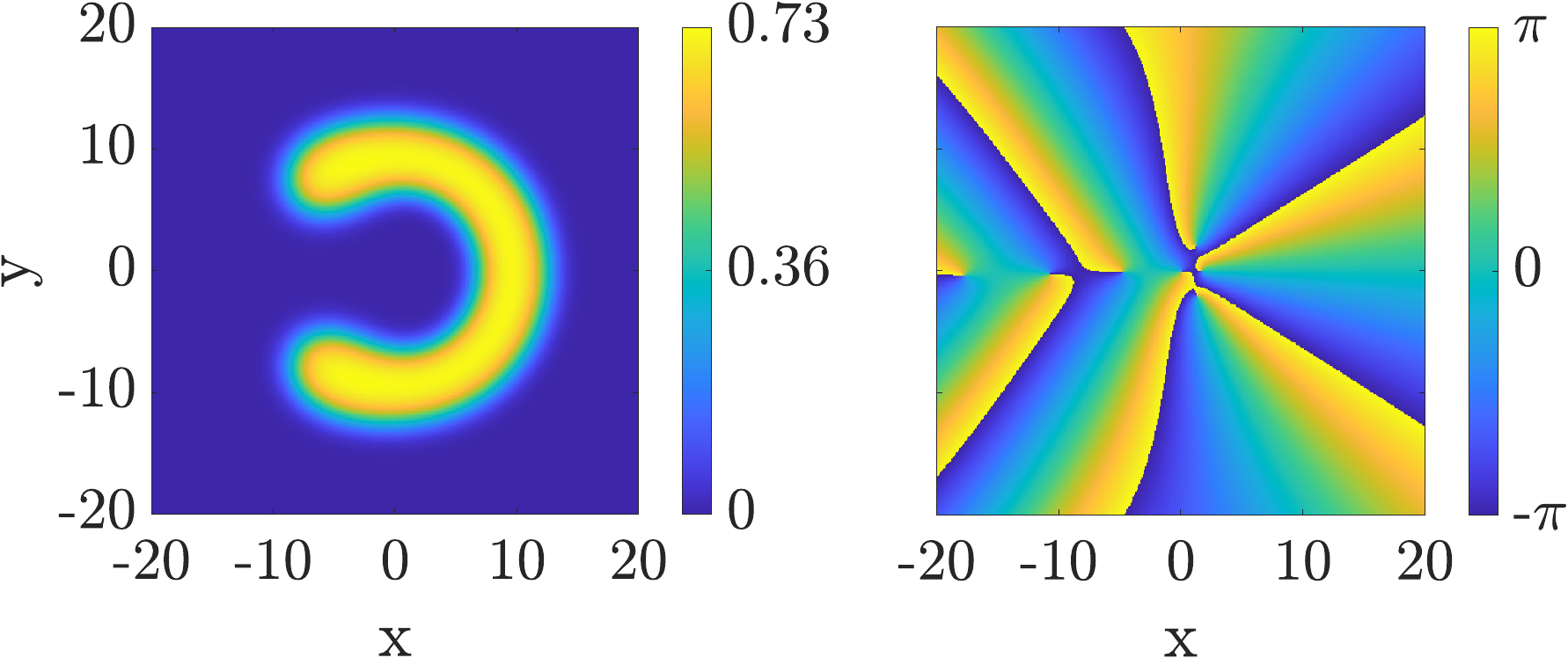}
\caption{(Cont.) The density (left column) and the phase 
(right column) of the droplet order parameter, for $N = 200$, $\omega = 0.05$, 
$\lambda = 0.05$, and $\ell = 3.0$, $3.5$, $4.0$, $4.5$, $5.0$, and $5.5$, from top to bottom. 
Here the density is measured in units of $\Psi_0^2$ and the length in units of 
$x_0$.}
\end{figure}

\begin{figure}
\includegraphics[width=\columnwidth ,angle=-0]{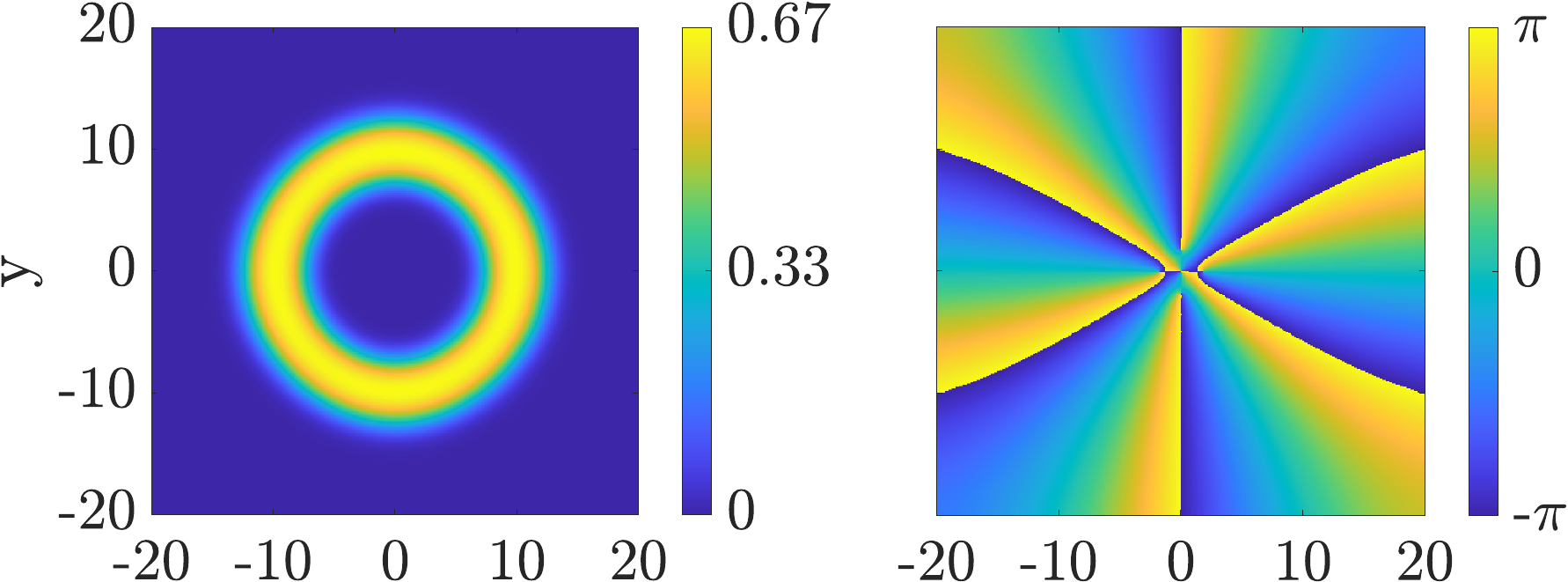}\\
\vspace{0.3\baselineskip}
\includegraphics[width=\columnwidth ,angle=-0]{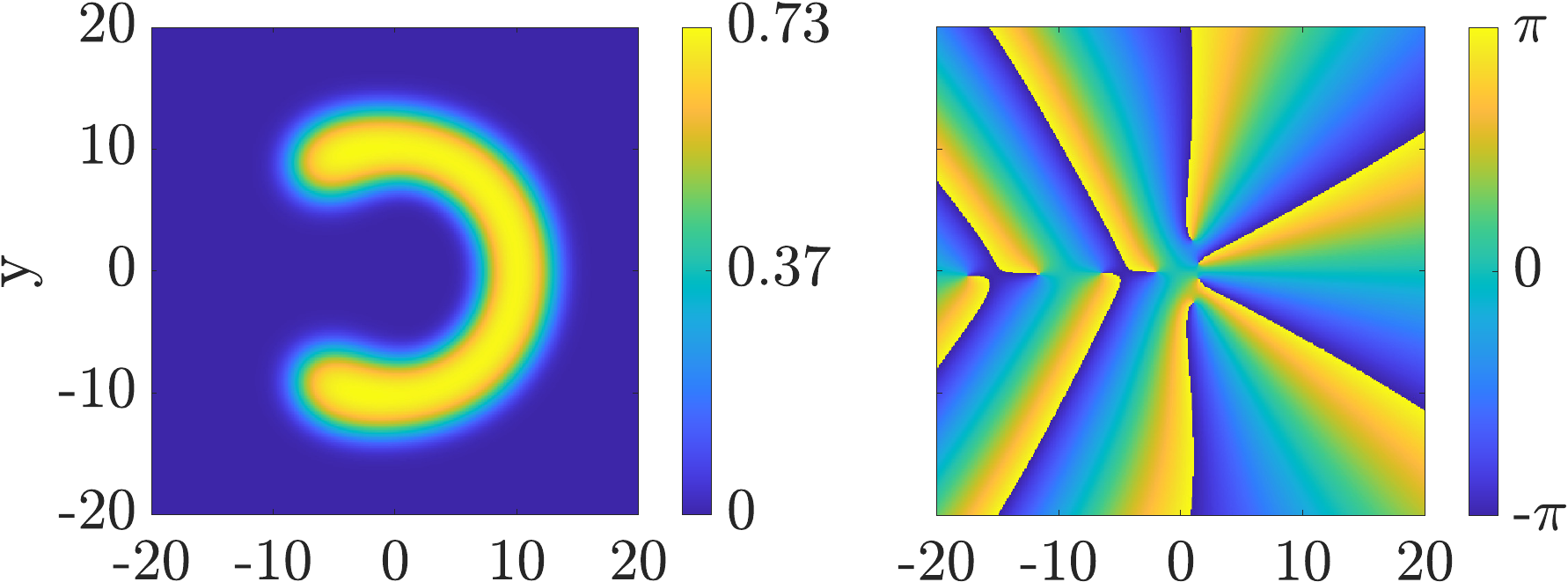}\\
\vspace{0.3\baselineskip}
\includegraphics[width=\columnwidth ,angle=-0]{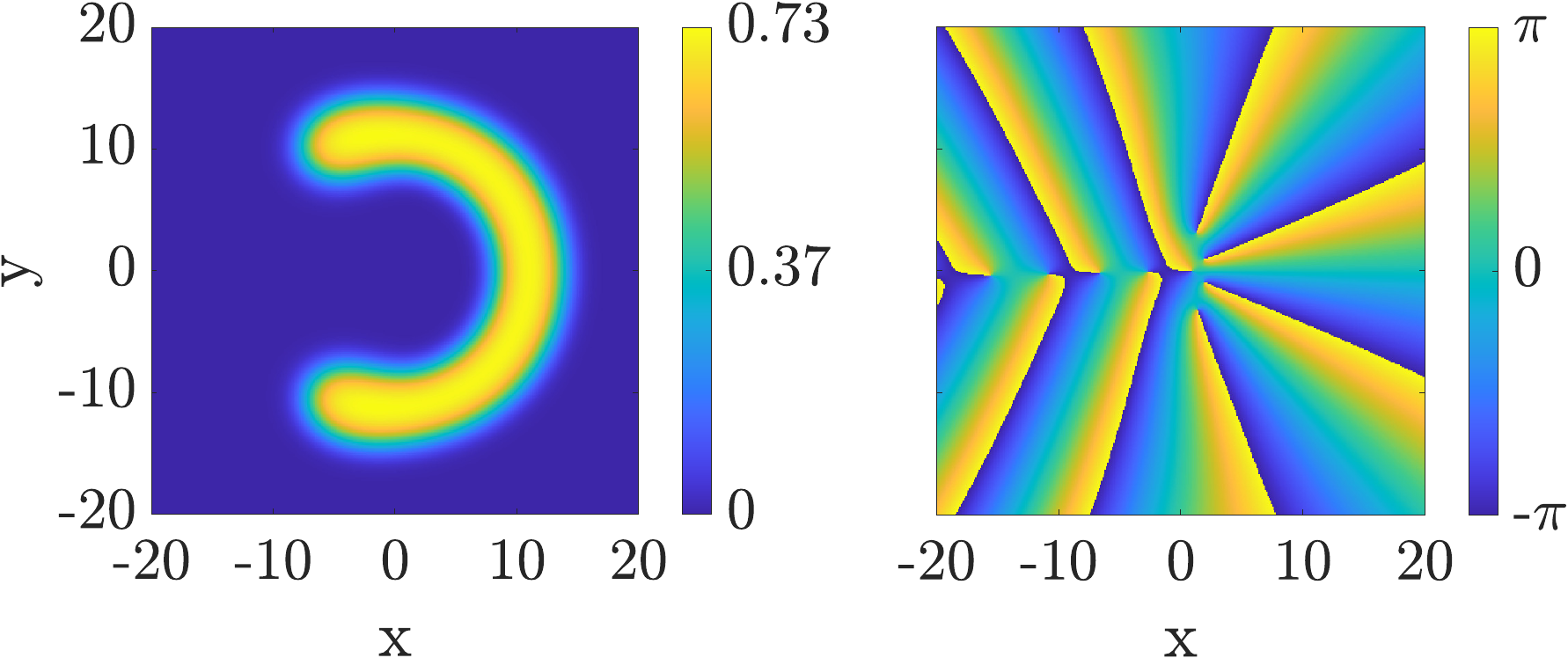}\\
\vspace{\baselineskip}
\includegraphics[width=\columnwidth ,angle=-0]{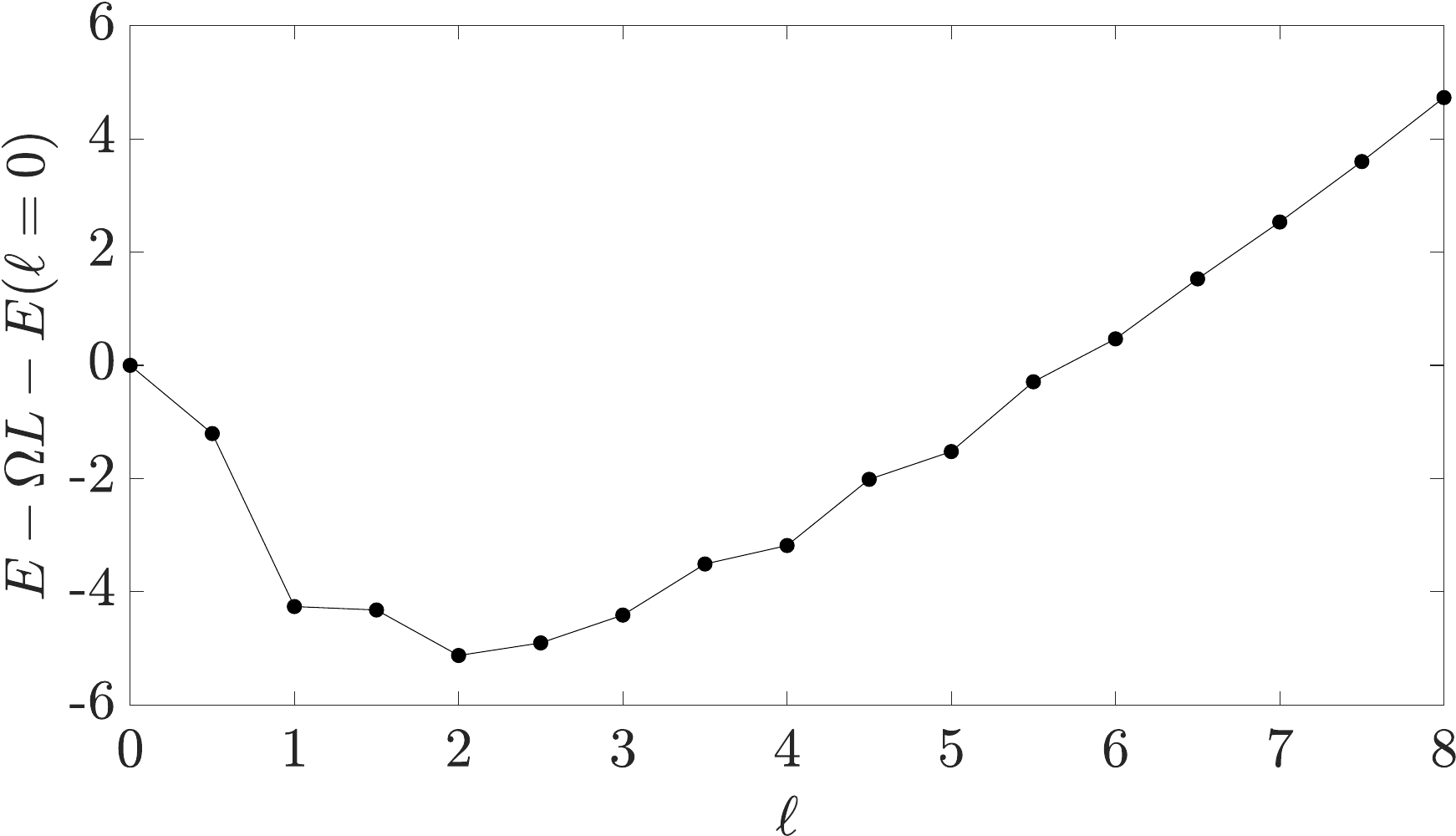}
\caption{Upper plots: The density (left column) and the phase 
(right column) of the droplet order parameter, for $N = 200$, $\omega = 0.05$, 
$\lambda = 0.05$, and $\ell = 6.0$, $6.5$, and $8.0$, from top to bottom. 
Here the density is measured in units of $\Psi_0^2$ and the length in units of 
$x_0$. Lower plot: The corresponding dispersion relation, in the rotating 
frame, i.e., $E_{\rm rot}(\ell) - E(\ell = 0)$ as function of $\ell$, with $\Omega 
= 0.054$. Here the energy is measured in units of $E_0$ and the angular momentum 
in units of $\hbar$.}
\end{figure}

The energy $E(\ell)$ shown at the bottom of Fig.\,4 in the rotating frame for
$\Omega = 0.054$, develops an even more interesting structure as
compared to Figs.\,1 and 2. Depending on the value of $\Omega$, this function has 
minima for various values of $\ell$.

\subsection{Fixing $\Omega$ instead of $L$}

Up to now, all the calculations that we have performed were for fixed angular 
momentum. From the derived dispersion relation $E(\ell)$ one may also see how the 
droplet would respond if $\Omega$ is fixed, instead. This is done by considering 
the energy in the rotating frame, i.e., $E_{\rm rot} = E(\ell) - \Omega L$, and 
locating the minimum. This is how Fig.\,6 is derived, where we show the
function $\ell = \ell(\Omega)$, for $N = 50$, $100$, $150$, and $200$. 

\begin{figure}[ht!]
\includegraphics[width=\columnwidth ,angle=-0]{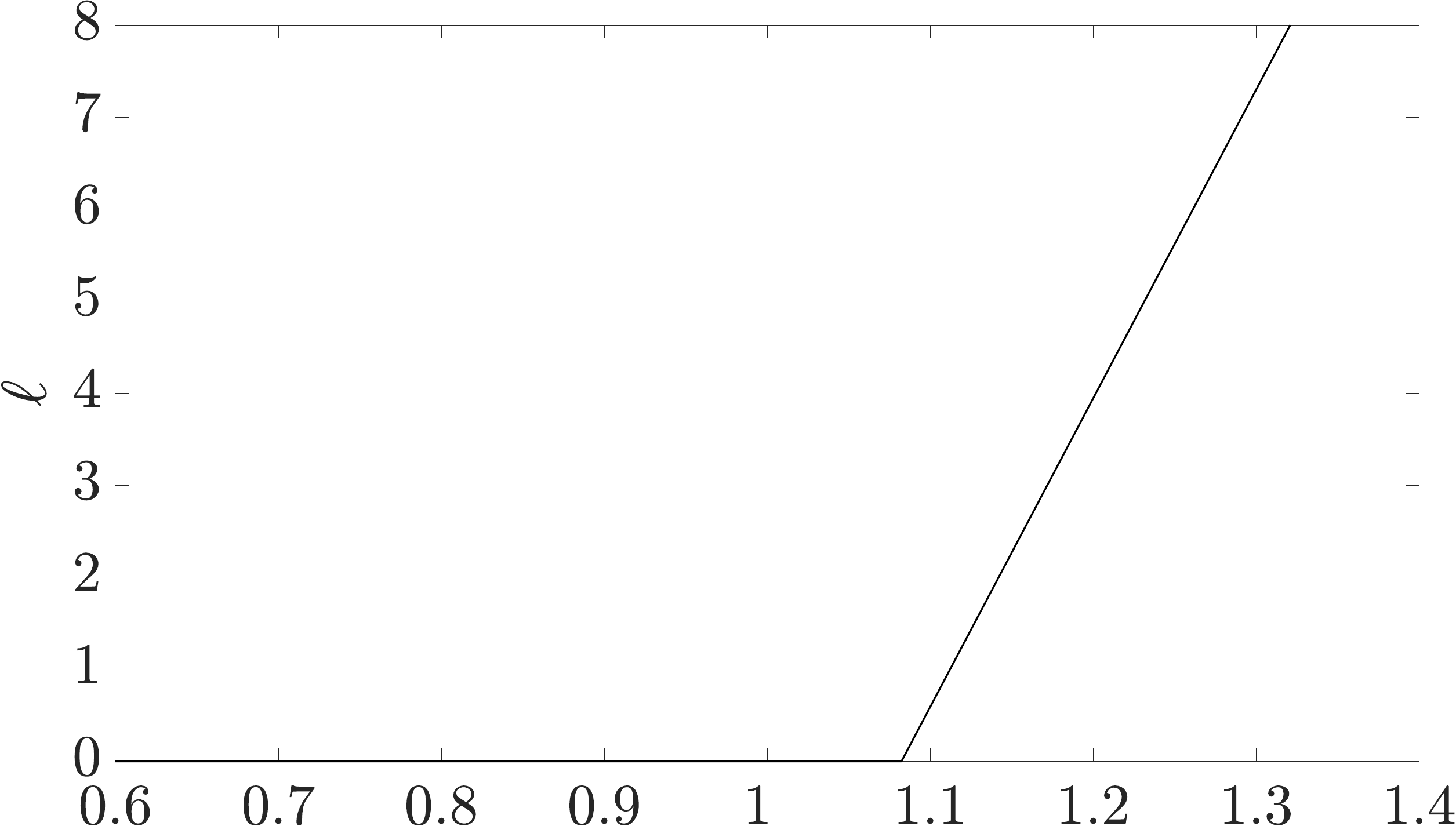}\\
\vspace{0.3\baselineskip}
\includegraphics[width=\columnwidth ,angle=-0]{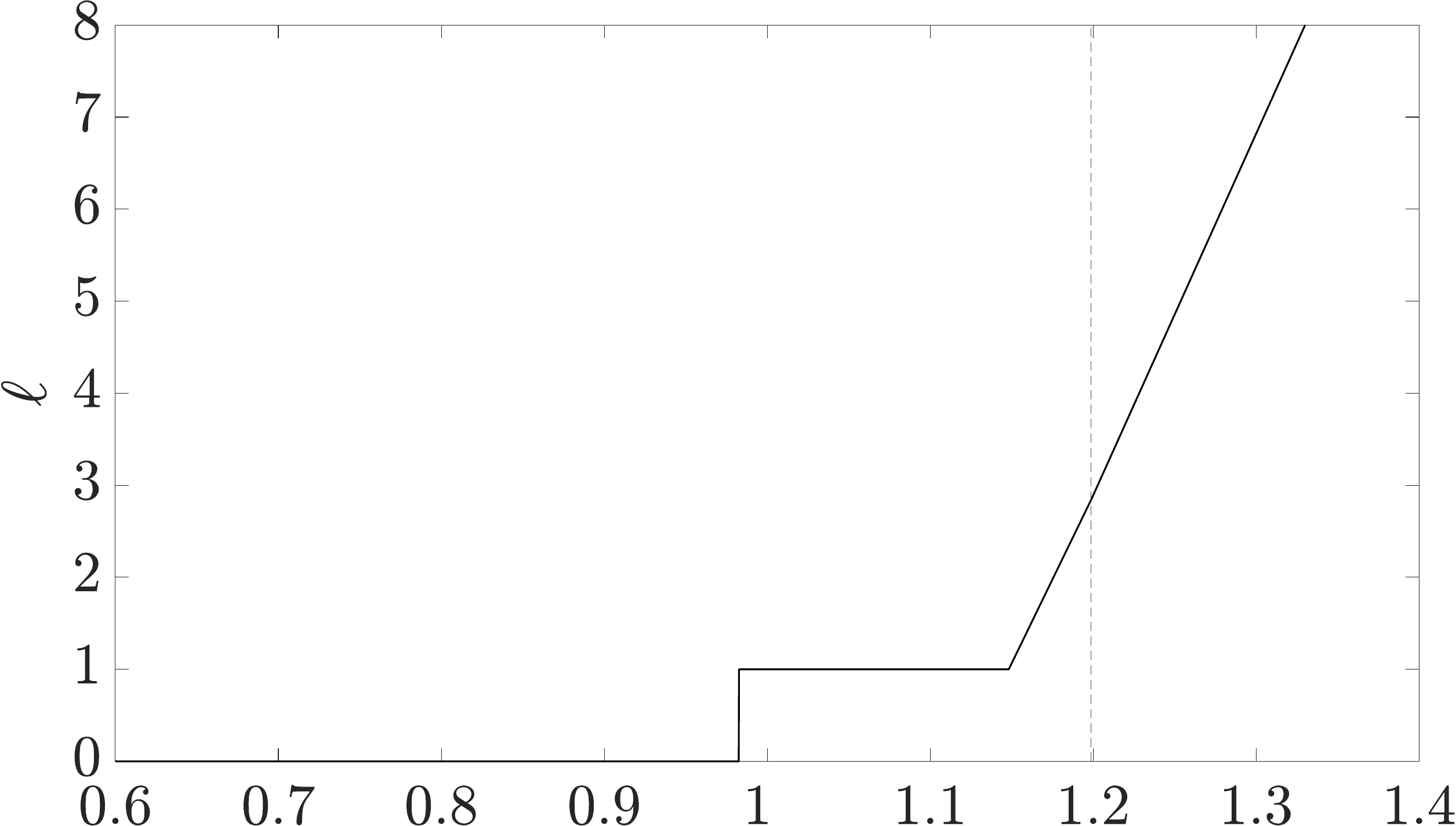}\\
\vspace{0.3\baselineskip}
\includegraphics[width=\columnwidth ,angle=-0]{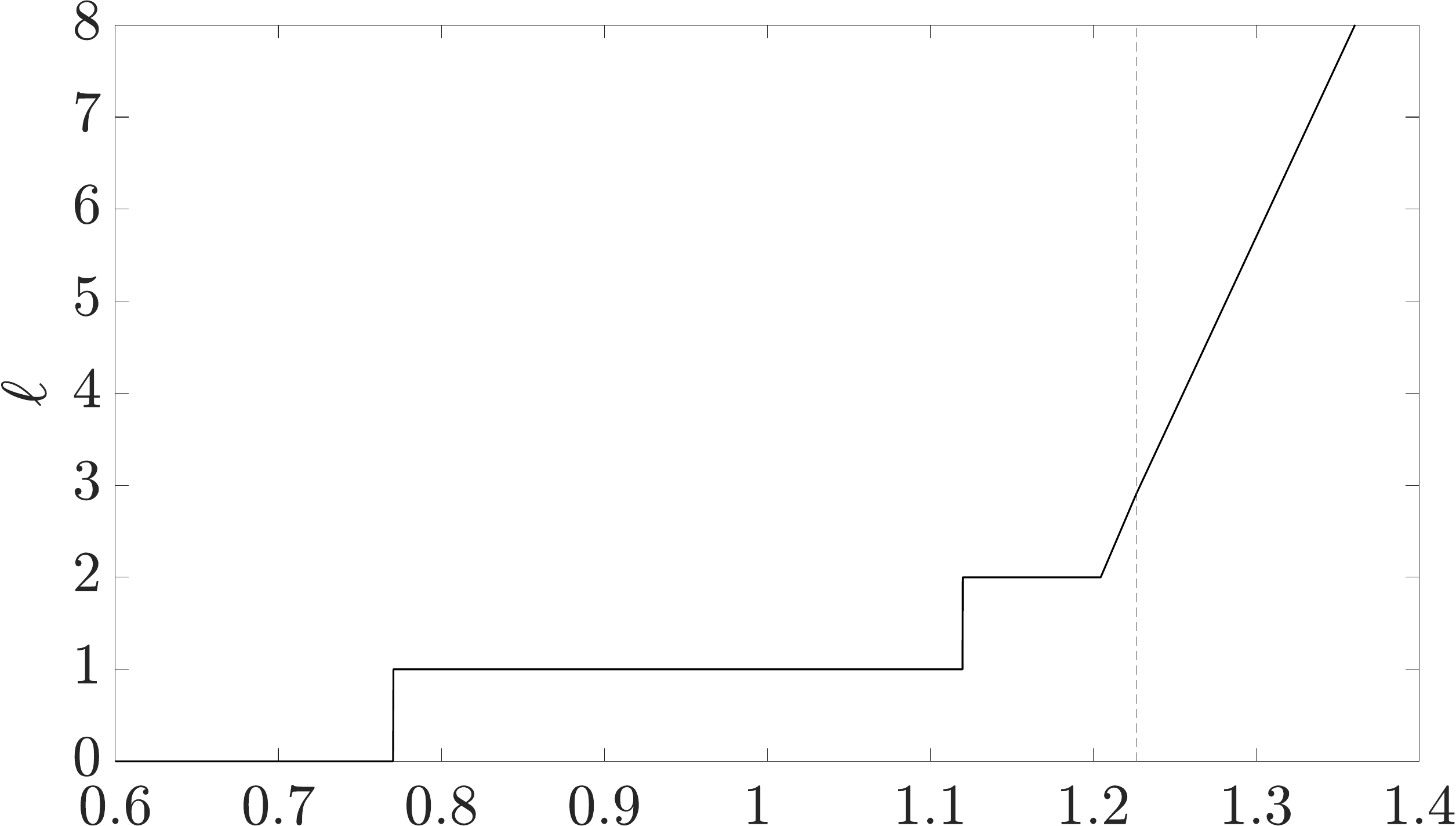}\\
\vspace{0.3\baselineskip}
\includegraphics[width=\columnwidth ,angle=-0]{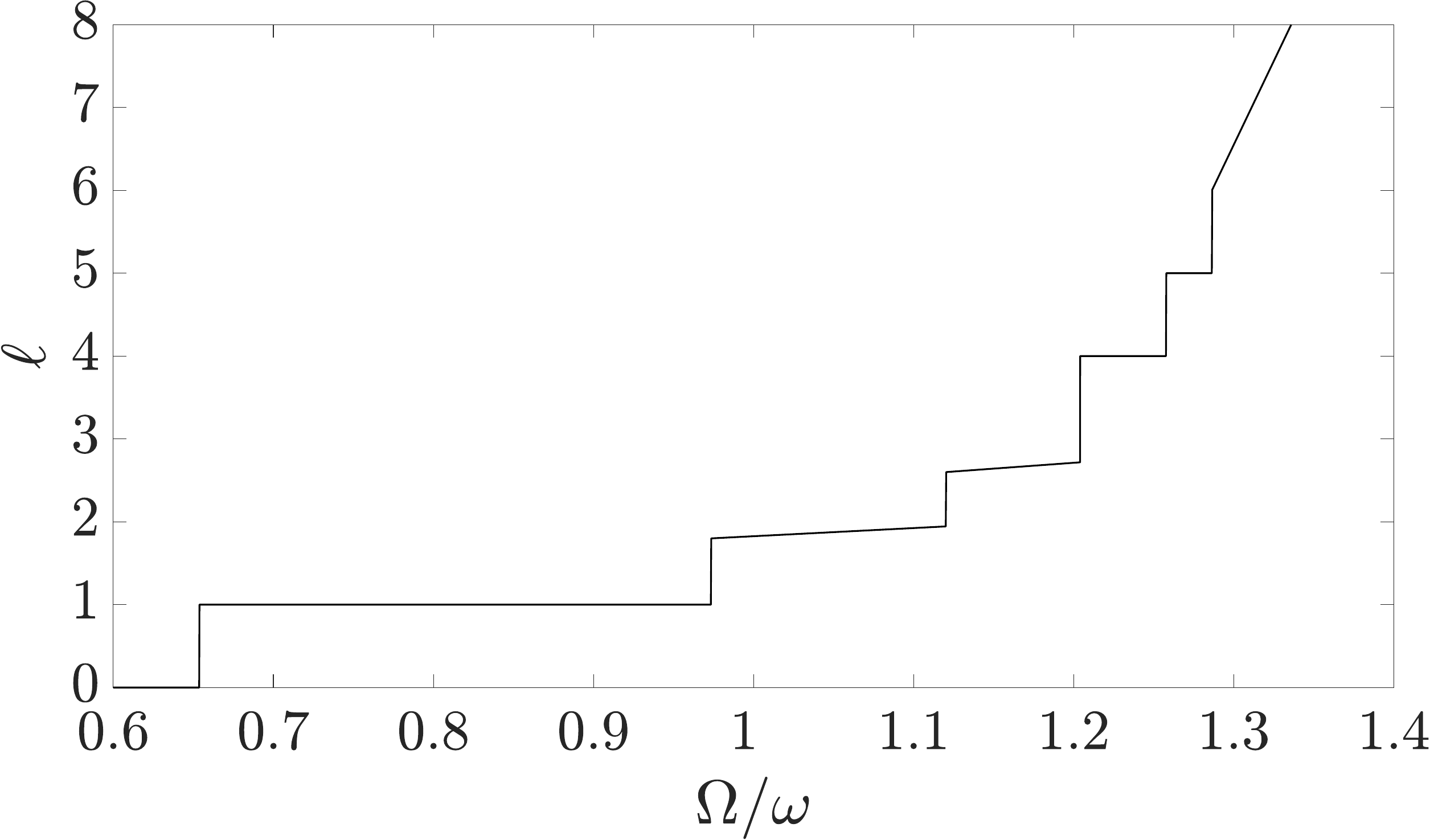}
\caption{The function $\ell = \ell(\Omega)$, for four values of $N = 50$, $100$, $150$, and $200$, from top to bottom, that we have considered. Dashed vertical lines denote a transition from a ``mixed" state to a localized state (see text). Here the angular momentum is measured in units of $\hbar$.}
\end{figure}

For the smallest value, $N = 50$, we saw that the droplet undergoes
center-of-mass-like excitation (see Fig.\,1). As a result, for fixed $\Omega$, 
there is a critical value of this frequency below which the droplet does not 
respond and is static. This critical value is $\Omega \approx 0.054117$, i.e., it is the coefficient of the linear term in Eq.\,(\ref{quadfit}), as expected. When $\Omega$ exceeds this value, the angular momentum 
starts to increase linearly with $\Omega$, according to the classical expression 
$L = I \Omega$, where $I$ is the moment of inertia of the droplet.

For $N = 100$ the picture changes (see Fig.\,2), as is seen in Fig.\,6. In this 
case, there is again a critical value of $\Omega$ below which the droplet is static, 
however when $\Omega$ exceeds this value, the droplet undergoes a discontinuous (due to the negative curvature of the dispersion relation for $0 < \ell <1$) transition to a state with a vortex that is located at its center. As $\Omega$ 
increases further, the droplet undergoes a transition to a ``mixed" state, where $\ell = \ell(\Omega)$ is a linear function. Finally the droplet transitions to a localized state, where we have center-of-mass-like excitation. Again, in this case $\ell = \ell(\Omega)$ is a linear function, according to the classical formula $L = I \Omega$. For $N = 150$ the picture is essentially similar (see Fig.\,3), with the addition of an extra step in $\ell = \ell(\Omega)$, located at $\ell = 2$.

It is important to note here that in the $\ell = \ell(\Omega)$ plot, $N = 100$ and $150$ each manifests two linear regions (i.e., with different slopes), one corresponding to ``mixed" excitation and one corresponding to center-of-mass-like excitation. However, we stress that the difference among these two slopes, for each value of $N$, is minuscule. As a result, the linear regions for $N = 100$ and $150$ appear to be uniform in Fig.\,6.

Finally, for $N = 200$, a richer picture, as compared to $N = 100$ and $150$, emerges (see Figs.\,4 and 5). 
One difference is that the critical value of $\Omega$ for the entry of the first 
vortex decreases. Another difference is that there are more steps in $\ell = 
\ell(\Omega)$, before the droplet (again) gets to a localized state, where (again) 
we have center-of-mass-like excitation. Here, there is no linear part corresponding to ``mixed" excitation, i.e., no ``mixed" state appears as an energy minimum in the rotating frame. The formula $L = I \Omega$ also holds here. 

A general observation about the center-of-mass-like excitation is that, while the 
formula $L = I \Omega$ is always valid, with increasing $N$, the value of $I$ increases, 
too. As a result, the slope of the linear part of the plotted functions increases as $N$ 
increases, as is seen clearly in Fig.\,6.

An interesting observation regards the order of transition to the ``mixed" states. For $N = 100$, the transition from the singly-quantized vortex state to the ``mixed" state is continuous. Conversely, for $N = 150$, the transition from the doubly-quantized vortex state to the ``mixed" state is discontinuous, i.e., first-order \footnote{Here, there also exists a continuous transition, from the doubly-quantized vortex state to a ``mixed" state, with center-of-mass-like excitation of the doubly-quantized vortex. However, this particular ``mixed" state is an excited state, similar to the excited states identified in Ref.\,\cite{NKO}, but with an axially-asymmetric density distribution.}. The discontinuity here arises from the negative curvature of the dispersion relation for $2 < \ell \lesssim 3$. We stress that, when needed, we have used a step in the $\ell$ values that is smaller than $0.5$ (which is used in the plots of the dispersion relations), down to $0.01$, so that we can accurately determine the curvature, and therefore the order of transitions.

\begin{figure}
\includegraphics[width=\columnwidth ,angle=-0]{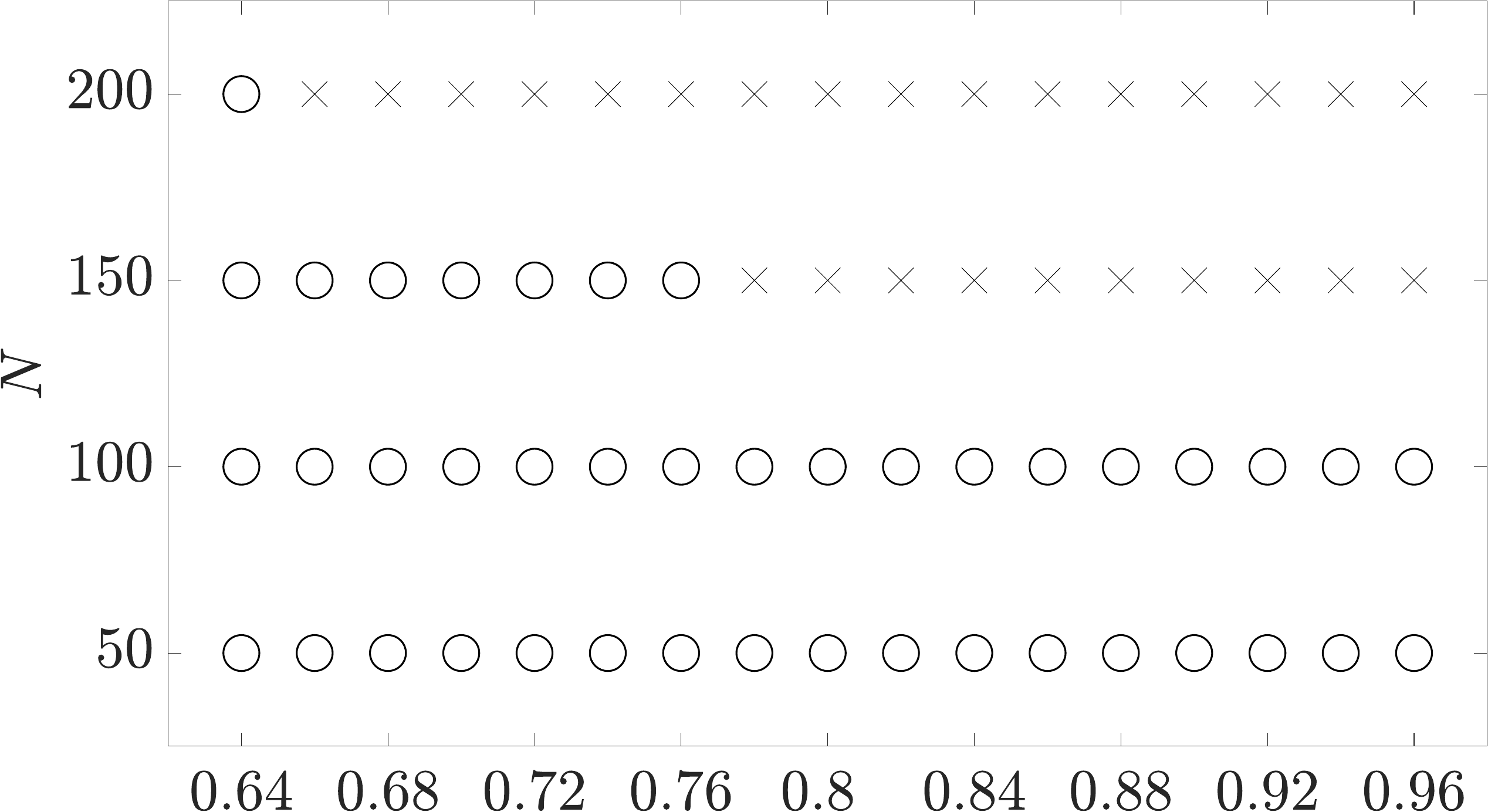}\\
\vspace{0.3\baselineskip}
\includegraphics[width=\columnwidth ,angle=-0]{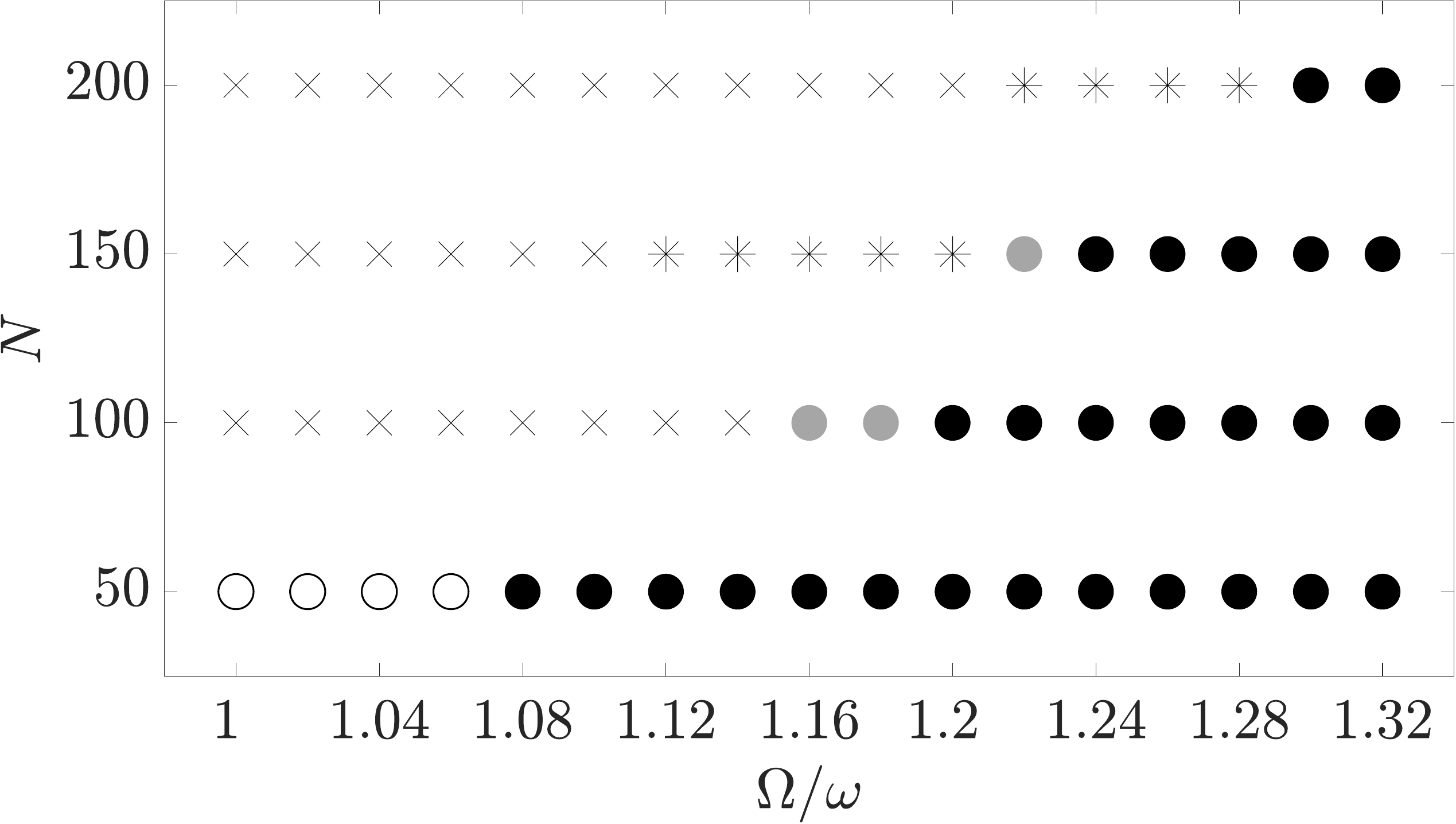}
\caption{The various phases that we have derived, on the $\Omega/\omega$ -- $N$ 
plane, which correspond to the absolute minima of the energy in the rotating frame, 
see Sec.\,III E, for ``slow" (top) and ``rapid" (bottom) rotation. On the horizontal axis 
is $\Omega/\omega$ and on the vertical is $N$. Here hollow circles denote the non-rotating 
ground state, crosses denote vortices of single quantization/vortex lattices,
asterisks denote vortices of multiple quantization, gray, solid circles, denote the 
``mixed" phase (see text), and black, solid circles, denote the center-of-mass-like, 
localized state.}
\end{figure}

We can also comment on the order of transition to the localized states with center-of-mass-like excitation. For $N=50$, the transition from the static droplet to center-of-mass-like excitation is continuous. However, for $N=100$, $150$ and $200$ the transition to a localized state is discontinuous. In particular, for $N=100$ and $150$ there is a level-crossing between the branch of the ``mixed" states and the branch of center-of-mass-like excitation. This level-crossing is located at $\Omega \approx 0.05994$ for $N=100$, and at $\Omega \approx 0.06134$ for $N=150$, corresponding to the dashed vertical lines in Fig.\,6.

\section{General picture -- Phase diagram}

From the results that are presented in the previous sections, it is clear that 
the problem we have considered has a very rich structure. In this section we 
give some general features of the phase diagram that includes the rotational 
frequency of the trap on the one axis and the atom number on the other axis, 
concentrating on the states which minimize the energy in the rotating frame.

As mentioned also earlier, we have considered the case where both the harmonic, 
as well as the anharmonic terms in the energy are smaller than the energy that results 
from the nonlinear term. In the opposite limit the droplet is ``squeezed" by the trap 
and the physics is -- at least qualitatively -- similar to the one-component system, 
with an effective repulsive interaction. This is due to the fact that when the density 
exceeds (sufficiently) the density of the droplet in free space, the nonlinear term 
becomes (predominantly) positive.

Figure 7 shows the phase diagram, where the data correspond to the ones presented 
in Sec.\,III E. For a sufficiently small atom 
number $N$, the only phase that is present, for all values of $\Omega$, is the 
one which resembles center-of-mass excitation. This is due to the fact that for 
the assumed small value of $N$ the droplet size is also ``small" and is not 
affected by the presence of the trapping potential. 

As $N$ increases (i.e., as we move vertically in the phase diagram), the droplet 
expands radially and starts to get ``squeezed" by the external trapping potential. 
As a result, the nonlinear term becomes predominantly repulsive. Furthermore, the 
energy due to both the harmonic and the anharmonic parts of the trapping potential 
increases. As a result, the system no longer undergoes center-of-mass-like excitation, 
but rather it supports vortex states, either of multiple, or of single quantization 
(for even larger values of $N$). Although not presented in this work, we have even identified another ``mixed" phase, for sufficiently large values of $N$, which contains a hole (i.e., a multiply-quantized vortex) at the center of the droplet, and singly-quantized vortices around it. This phase has also been identified in anharmonically-confined, rotating Bose-Einstein condensates with effectively-repulsive contact interactions \cite{qq8}.

For a fixed atom number and with increasing $\Omega$ (i.e., as we move horizontally 
in the phase diagram), the physics of a droplet is determined by the effective 
potential, which has a Mexican-hat shape for $\Omega > \omega$. Then, the droplet 
forms either vortices of multiple quantization, or a localized blob around the minimum 
of the effective potential, in a state which breaks the axial symmetry of the 
Hamiltonian. 

\section{Experimental relevance}

So far we have been working with dimensionless units for convenience.
Here we give some numbers which relate with the physical units and allow one to make
a connection with actual experiments. 

For a typical value of $a_z = 0.1$ $\mu$m and $a^{\rm 3D} = 10.1$ nm, $a_{\uparrow 
\downarrow}^{\rm 3D} = -10.0$ nm, $\ln (a_{\uparrow \downarrow}/a) \approx 25$. Then, 
according to Eq.\,(\ref{n0}), $N_0 \approx 50$. Therefore, the range of $N$ that we 
have considered (50 up to 200) corresponds roughly to $\approx 2500$, up to $\approx 
10000$ atoms in an experiment.

Also, the unit of length $x_0$ turns out to be on the order of 1 $\mu$m.
This implies that, for e.g., $10^4$ atoms, the size of a (non-rotating) droplet 
in the Thomas-Fermi limit, which was evaluated in Sec.\,III, is $\approx 10$ $\mu$m.
A typical value of the two-dimensional density is $\approx 10^9$ ${\rm cm}^{-2}$, 
of the three-dimensional density is $10^{13}$ ${\rm cm}^{-3}$, the unit of time $t_0$ is on the order of millisecond and the typical value of $\omega$ is hundreds of hertz. Finally, a 
typical value of the anharmonicity parameter $\lambda$ is $\approx 10^{-2}$ \cite{Dalibard}. \bigskip

\section{Summary}

In the present manuscript we investigated the rotational
properties of a mixture of two Bose-Einstein condensates, which consists 
of an equal population of distinguishable atoms and also have an equal mass. 
Under the further assumption that the mean-field energy of this binary
mixture is sufficiently small, the next-order correction to the energy 
is non-negligible (as opposed to most other cases) and the balance between 
the two terms results in the formation of quantum droplets.

The presence of an (even weak) anharmonic term in the potential has very 
serious consequences on the rotational response of the gas. First of all, 
for a fixed rotational frequency $\Omega$, while in a harmonic potential 
$\Omega$ cannot exceed $\omega$, here there is no such restriction.
Furthermore, while in a harmonic potential the center-of-mass coordinate 
separates from the relative coordinates, here this is no longer true.

Given that droplets are self-bound states, for a sufficiently weak trapping 
potential and/or a sufficiently small atom number $N$, the nonlinear term 
is attractive. The droplet then carries its angular momentum in a state that 
resembles center-of-mass excitation (with some distortion, though, since, 
as we mentioned in the previous paragraph the center-of-mass coordinate does 
not separate from the relative coordinates). As the trapping potential becomes 
stronger and/or the atom number $N$ increases, the nonlinear term becomes 
(predominantly) repulsive. In this case, it becomes energetically favourable 
for the droplet to accommodate vortices, either of multiple, or of single 
quantization. 

The above results are in contrast to the case of a single component, where in 
the limit of a weak trapping potential and/or a sufficiently small atom number 
$N$, one always has vortices of multiple quantization, (and this is actually 
true for both signs of the effective interaction).

For a fixed atom number and with increasing $\Omega$, the droplet forms either 
vortices of multiple quantization, or a localized blob around the minimum 
of the effective potential, in a state which breaks the axial symmetry of the 
Hamiltonian (depending on the actual value of the atom number).

We conclude by stressing that, given that in actual experiments there are 
always deviations from a perfectly harmonic trap, the assumption of an
anharmonic potential is probably more realistic and more experimentally-relevant 
as compared to a model of rotating quantum droplets in a purely harmonic potential. Clearly, this adds
more value to the results of the present study.


\begin{thebibliography}{55}

\bibitem{Leggett} A. J. Leggett, Rev. Mod. Phys. {\bf 71}, S318 (1999).

\bibitem{Cooper} N. R. Cooper, Adv. Phys. {\bf 57}, 539 (2008).

\bibitem{Dalibard} Vincent Bretin, Sabine Stock, Yannick Seurin, and Jean
Dalibard, Phys. Rev. Lett. {\bf 92}, 050403 (2004).

\bibitem{qq1} A. L. Fetter, Phys. Rev. A {\bf 64}, 063608 (2001).

\bibitem{qq2} E. Lundh, Phys. Rev. A {\bf 65}, 043604 (2002).

\bibitem{qq3} K. Kasamatsu, M. Tsubota, and M. Ueda, Phys. Rev. A 
{\bf 66}, 053606 (2002).

\bibitem{qq4} U. R. Fischer and G. Baym, Phys. Rev. Lett. {\bf 90}, 140402
(2003).

\bibitem{qq5} G. M. Kavoulakis and G. Baym, New J. Phys. {\bf 5}, 51 (2003).

\bibitem{qq6} E. Lundh, A. Collin, and K.-A. Suominen, Phys. Rev. Lett. {\bf 92},
070401 (2004).

\bibitem{qq7} A. Aftalion and I. Danaila, Phys. Rev. A {\bf 69}, 033608 (2004).

\bibitem{qq8} A. D. Jackson, G. M. Kavoulakis, and E. Lundh, Phys. Rev. A {\bf 69}, 
053619 (2004).

\bibitem{qq9} Alexander L. Fetter, B. Jackson, and S. Stringari, Phys. Rev. A {\bf 71}, 
013605 (2005).

\bibitem{Petrov} D. S. Petrov, Phys. Rev. Lett. {\bf 115}, 155302 (2015).

\bibitem{LHY} T. D. Lee, K. Huang, and C. N. Yang, Phys. Rev. {\bf 106}, 1135 (1957).
 
\bibitem{rrev1}F. B\"ottcher, J.-N. Schmidt, J. Hertkorn, K. S. H. Ng, S. D. Graham,
M. Guo, T. Langen, and T. Pfau, Reports on Progress in Physics {\bf 84}, 012403 
(2021).

\bibitem{rrev2} Z.-H. Luo, W. Pang, B. Liu, Y.-Y. Li, and B. A. Malomed,
Frontiers of Physics {\bf 16}, 32201 (2021).

\bibitem{PA} D. S. Petrov and G. E. Astrakharchik, Phys. Rev. Lett. {\bf 117},
100401 (2016).

\bibitem{th0} Yongyao Li, Zhihuan Luo, Yan Liu, Zhaopin Chen, Chunqing Huang, Shenhe Fu, 
Haishu Tan, and Boris A. Malomed, New J. Phys. {\bf 19}, 113043 (2017).

\bibitem{th1} G. E. Astrakharchik and B. A. Malomed, Phys. Rev. A {\bf 98}, 013631 
(2018). 

\bibitem{th2} Y. V. Kartashov, B. A. Malomed, L. Tarruell, and L. Torner, Phys. Rev.
A {\bf 98}, 013612 (2018).

\bibitem{th3} A. Cidrim, F. E. A. dos Santos, E. A. L. Henn, and T. Macr\'i, Phys.
Rev. A {\bf 98}, 023618 (2018).

\bibitem{th4} Pawe\l{} Zin, Maciej Pylak, Tomasz Wasak, Mariusz Gajda, and Zbigniew 
Idziaszek, Phys. Rev. A {\bf 98}, 051603(R) (2018). 

\bibitem{th5} F. Ancilotto, M. Barranco, M. Guilleumas, and M. Pi, Phys. Rev. A {\bf
98}, 053623 (2018).

\bibitem{th6} Y. Li, Z. Chen, Z. Luo, C. Huang, H. Tan, W. Pang, and B. A. Malomed,
Phys. Rev. A {\bf 98}, 063602 (2018).

\bibitem{th7} L. Parisi, G. E. Astrakharchik, and S. Giorgini, Phys. Rev. Lett. {\bf 122}, 
105302 (2019). 

\bibitem{th8} Y. V. Kartashov, B. A. Malomed, and L. Torner, Phys. Rev. Lett. {\bf
122}, 193902 (2019).

\bibitem{th9} Xiliang Zhang, Xiaoxi Xu, Yiyin Zheng, Zhaopin Chen, Bin Liu, Chunqing 
Huang, Boris A. Malomed, and Yongyao Li, Phys. Rev. Lett. {\bf 123}, 133901 (2019).

\bibitem{th10} M. Nilsson Tengstrand, P. St\"urmer, E. \"O. Karabulut, and S. M. Reimann,
Phys. Rev. Lett. {\bf 123}, 160405 (2019). 

\bibitem{th11} Bin Liu, Hua-Feng Zhang, Rong-Xuan Zhong, Xi-Liang Zhang, Xi-Zhou Qin, 
Chunqing Huang, Yong-Yao Li, and Boris A. Malomed, Phys. Rev. A {\bf 99}, 053602 (2019). 

\bibitem{th12} R. Tamil Thiruvalluvar, S. Sabari, K. Porsezian, P. Muruganandam, 
Physica E {\bf 107}, 54 (2019). 

\bibitem{th13} G. Ferioli, G. Semeghini, S. Terradas-Brians\'{o}, L. Masi, 
M. Fattori, and M. Modugno, Phys. Rev. Research {\bf 2}, 013269 (2020). 

\bibitem{th14} Ivan Morera, Grigori E. Astrakharchik, Artur Polls, and Bruno 
Juli\'{a}-D\'{i}az, Phys. Rev. Research {\bf 2}, 022008(R) (2020). 

\bibitem{th15} Luca Parisi and Stephano Giorgini, Phys. Rev. A {\bf 102}, 023318 (2020). 

\bibitem{th16} Marek Tylutki, Grigori E. Astrakharchik, Boris A. Malomed, and 
Dmitry S. Petrov, e-print arXiv:2003.05803. 

\bibitem{EK} P. Examilioti and G. M. Kavoulakis, J. Phys. B: At. Mol. Opt. Phys. 
{\bf 53}, 175301 (2020).

\bibitem{th166} Yanming Hu, Yifan Fei, Xiao-Long Chen and Yunbo Zhang, 
Frontiers of Physics {\bf 17}, 61505 (2022). 

\bibitem{add1} Szu-Cheng Cheng, Yu-Wen Wang, and Wen-Hsuan Kuan,
e-print arXiv:2302.07481.

\bibitem{NKO} S. Nikolaou, G. M. Kavoulakis, and M. {\"O}gren, Phys. Rev. A {\bf 108}, 053309 (2023). 

\bibitem{add2} Qi Gu and Xiaoling Cui, e-print arXiv:2306.14958.

\bibitem{add3} T. A. Yo\u{g}urt, U. Tanyeri, A. Kele\c{s}, and M. \"O. Oktel, Phys. Rev. A {\bf 108}, 033315 (2023).

\bibitem{add4} T. A. Flynn, N. A. Keepfer, N. G. Parker, and T. P. Billam, e-print arXiv:2309.04300.

\bibitem{add5} Xucong Du, Yifan Fei, Xiao-Long Chen, and Yunbo Zhang, e-print arXiv:2309.05245.

\bibitem{qd7} C. Cabrera, L. Tanzi, J. Sanz, B. Naylor, P. Thomas, P. Cheiney, and
L. Tarruell, Science {\bf 359}, 301 (2018).

\bibitem{qd8} P. Cheiney, C. R. Cabrera, J. Sanz, B. Naylor, L. Tanzi, and L. Tarruell, 
Phys. Rev. Lett. {\bf 120}, 135301 (2018).

\bibitem{qd8a} G. Semeghini, G. Ferioli, L. Masi, C. Mazzinghi, L. Wolswijk, F.
Minardi, M. Modugno, G. Modugno, M. Inguscio, and M. Fattori, Phys. Rev. Lett. {\bf 120}, 
235301 (2018).

\bibitem{gd8b} Giovanni Ferioli, Giulia Semeghini, Leonardo Masi, Giovanni Giusti, 
Giovanni Modugno, Massimo Inguscio, Albert Gallem\'{i}, Alessio Recati, and Marco 
Fattori, Phys. Rev. Lett. {\bf 122}, 090401 (2019).

\bibitem{qd8c} C. D'Errico, A. Burchianti, M. Prevedelli, L. Salasnich, F. Ancilotto, 
M. Modugno, F. Minardi, and C. Fort, Phys. Rev. Research {\bf 1}, 033155 (2019).

\bibitem{qd1} H. Kadau, M. Schmitt, M. Wenzel, C.Wink, T. Maier, I. Ferrier-Barbut,
and T. Pfau, Nature {\bf 530}, 194 (2016).

\bibitem{qd2} M. Schmitt, M. Wenzel, F. B\"ottcher, I. Ferrier-Barbut, and T. Pfau,
Nature {\bf 539}, 259 (2016).

\bibitem{qd3} I. Ferrier-Barbut, H. Kadau, M. Schmitt, M. Wenzel, and T. Pfau, Phys.
Rev. Lett. {\bf 116}, 215301 (2016).

\bibitem{qd4} I. Ferrier-Barbut, M. Schmitt, M. Wenzel, H. Kadau, and T. Pfau, J.
Phys. B {\bf 49}, 214004 (2016).

\bibitem{qd5} I. Ferrier-Barbut, M. Wenzel, F. B\"ottcher, T. Langen, M. Isoard, S.
Stringari, and T. Pfau, Phys. Rev. Lett. {\bf 120}, 160402 (2018).

\bibitem{qd6} L. Chomaz, S. Baier, D. Petter, M. J. Mark, F. W\"achtler, L. Santos,
and F. Ferlaino, Phys. Rev. X {\bf 6}, 041039 (2016).

\bibitem{rapanh} Liangwei Dong and Yaroslav V. Kartashov, Phys. Rev. Lett. {\bf 126}, 
244101 (2021).

\bibitem{dynamics} Szu-Cheng Cheng, Yu-Wen Wang, and Wen-Hsuan Kuan, e-print
arXiv:2302.07481.

\bibitem{gv1} A. L. Fetter, Phys. Rev. A {\bf 64}, 063608 (2001).

\bibitem{gv2} E. Lundh, Phys. Rev. A {\bf 65}, 043604 (2002).

\bibitem{gv3} K. Kasamatsu, M. Tsubota, and M. Ueda, Phys. Rev. A {\bf 66},
053606 (2002).

\bibitem{gv4} G. M. Kavoulakis and G. Baym, New J. Phys. {\bf 5}, 51.1 (2003).

\bibitem{gv5} U. R. Fischer and G. Baym, Phys. Rev. Lett. {\bf 90}, 140402 (2003).

\bibitem{gv6} A. Aftalion and I. Danaila, Phys. Rev. A {\bf 69}, 033608 (2004).

\bibitem{gv7} A. D. Jackson, G. M. Kavoulakis, and E. Lundh, Phys. Rev. A {\bf 69}, 
053619 (2004).

\bibitem{gv8} A. D. Jackson and G. M. Kavoulakis, Phys. Rev. A {\bf 70}, 023601 
(2004).

\bibitem{gv9} E. Lundh, A. Collin, and K.-A. Suominen, Phys. Rev. Lett. {\bf 92}, 
070401 (2004).

\bibitem{gv10} G. M. Kavoulakis, A. D. Jackson, and Gordon Baym, Phys. Rev. A
{\bf 70}, 043603 (2004).

\bibitem{gv11} Anssi Collin, Emil Lundh, and Kalle-Antti Suominen, Phys. Rev. A 
{\bf 71}, 023613 (2005).

\bibitem{Elife} E. \"O. Karabulut, F. Malet, G. M. Kavoulakis, and S. M. Reimann,
Phys. Rev. A {\bf 87}, 043609 (2013).

\bibitem{GO} M. Gulliksson and M. \"Ogren, J. Phys. A: Math. Theor. {\bf 54},
275304 (2021).

\end{thebibliography}
\end{document}